\documentclass[superscriptaddress,aip,numerical]{revtex4-1}

\usepackage{graphicx,amsmath,amssymb}
\usepackage{subfig}
\graphicspath{{./figs/}}

\usepackage{changebar,xcolor,ulem}
\newif\ifdiff
\difffalse
\ifdiff
  \newcommand{\removed}[1]{\removedfragile{#1}}
  \newcommand{\removedfragile}[1]{{\color{red}{\sout{#1}}}{}}
  \newcommand{\added}[1]{\addedfragile{#1}}
  
  \newcommand{\changed}[2]{\removed{#1}\added{#2}}
\else
  \newcommand{\removed}[1]{} 
  \newcommand{\removedfragile}[1]{}
  \newcommand{\added}[1]{#1}
  
  \newcommand{\changed}[2]{\added{#2}}
\fi

\begin{document}

\title{Influence of subgrid scale models on the buffer sublayer in channel flow}
\author{Liang Shi}
\email{gliang.shi@gmail.com}
\affiliation{Engineering Laboratory, National Institute of Standards and Technology, Gaithersburg, Maryland 20899, USA}
\date{\today}%

\begin{abstract}
Subgrid-scale (SGS) models are critical in large-eddy simulations (LES) 
of turbulent flows. In this paper we conduct a comparative study on different SGS models, 
including one-k-equation, wall-adapting local eddy-viscosity (WALE), Sigma and shear-constrained model. 
Wall-resolved LES simulations of channel flows are performed with a finite volume code 
at shear Reynolds number $Re_\tau = 395$.
In the simulations, the buffer sublayer turns out to be the most sensitive to the SGS model.
Through the analysis of the mean velocity, 
the second-order moments, the SGS viscosity and the fluctuating vorticities, it is shown that 
the WALE and Sigma model outperform others significantly in terms of fluctuations 
while the constrained model improves slightly the mean velocity. 
The results also indicate that the SGS dissipation influences strongly the velocity 
fluctuations but not the mean flow nor the log-layer mismatch.
\end{abstract}

\maketitle


\section{Introduction} \label{sec:intro}
The governing equations for LES simulaton of incompressible turbulent flows are the 
filtered Navier-Stokes equations~\cite{LesieurMetais_annRevFluidMech1996,MeneveauKatz_annRevFluidMech2000}, 
\begin{subequations}
  \begin{align}
    \nabla \cdot \widetilde{\textbf{u}} &=0, \\
    \partial_t \widetilde{\textbf{u}} + \widetilde{\textbf{u}}\nabla \widetilde{\textbf{u}} &= -\nabla 
    \widetilde{p} + \nu\nabla^2 \widetilde{\textbf{u}} - \nabla\cdot \tau,
  \end{align}
  \label{eq:nsEqHomoFilter}%
\end{subequations}
where $\widetilde{p}$ and $\widetilde{\textbf{u}}$ are the filtered pressure and velocity 
fields, and $\tau_{ij} = \widetilde{u_iu_j} - \widetilde{u_i}\widetilde{u_j}$ is the SGS stress 
tensor. The tilde denotes the filter operator which will be hereafter omitted for simplicity.
Modeling the SGS flow motions must be based on a functional form of 
$\tau$ using the filtered velocity $\textbf{u}$. 
This seemingly uncomplicated problem has been scrutinized for almost 
half century yet remains a challenge. 

The first SGS model was proposed in 1960s by Smagorinsky~\cite{Smagorinsky_MonWeatherRev1963}
and Lilly~\cite{Lilly_ibm1967}, and was inspired by the eddy-viscosity concept 
$\tau_{ij} = -2\nu_{SGS}S_{ij}$. Here $\nu_{SGS} = l_S^2|S|$ is the SGS viscosity, 
$S_{ij}=(u_{i,j}+u_{j,i})/2$ is the 
rate-of-strain tensor of the filtered velocity and $l_S=C_S\Delta$ denotes the Smagorinsky 
length scale. Even though it is affected by a consistent overprediction of velocity in the logarithmic 
region of boundary-layer flows (the log-layer mismatch problem), 
the Smagorinsky model is still widely used by CFD practitioners due to its simplicity and 
numerical stability. 
Many variants have been ever since developed to improve the Smagorinsky model. 
Schumann~\cite{Schumann_jcp1975} proposed a kinetic energy model where the SGS kinetic energy 
is solved by a transport equation and its square root 
is used as a velocity scale in the model. Another 
important development is the dynamic Smagorinsky model~\cite{GermanoCabot_pof1991}, which 
is based on the Germano identity and the scale-invariance assumption of the model coefficient 
near the cut-off scale. 
To account for situations where the scale-invariance assumption does not hold, e.g., 
near-surface regions in atmospheric boundary layer flows, 
Port\'e-Agel and Meneveau~\cite{PorteAgel_jfm2000} proposed a scale-dependent dynamic 
SGS model.
However, the averaging procedure in dynamic models renders difficult their application 
in complex flows. 
Another group of models is based on the invariants of a symmetric tensor that depends on 
the gradient of the filtered velocity~\cite{TriasOliva_pof2015}. This group includes,
among many others, the wall-adapting local eddy-viscosity model 
(WALE)~\cite{NicoudDucros_1999}, the Vreman's model~\cite{Vreman_pof2004}, the Verstappen's 
model~\cite{Verstappen_jsc2011} and the Sigma model~\cite{NicoudLee_pof2011}. 
Similar to the Smagorinsky model, these models are local and hence are 
suitable for complex flows and geometries. 

Another interesting idea on SGS modeling, which was also introduced by 
Schumann~\cite{Schumann_jcp1975}, is to decompose the SGS stress into a mean and a
fluctuating part following the concept of Reynolds' decomposition. 
The inhomogeneous effects near the wall, 
which cause problems in Smagorinsky model, are thus taken into account 
in the mean part. Based on this idea, many SGS models have been proposed, including 
the two-part model by Sullivan \textit{et al.}~\cite{SullivanMoeng_blm1994}, the shear-improved 
model by L\'ev\^eque et al.~\cite{LevequeBertoglio_jfm2007} and the recently-developed 
constrained SGS (CSGS) model~\cite{ChenShi_jfm2012}. 
The CSGS model uses the existing knowledge on the mean Reynold stress as 
a constraint in the SGS model and models only the fluctuation part. 
The knowledge of the mean quantity can be obtained from 
high-fidelity direct numerical simulation (DNS), experiments, analytical solutions or even low-fidelity RANS 
simulations. According to~\cite{ChenShi_jfm2012}, this model achieves accurate 
turbulent statistics while keeping the computational cost reasonable. Note that 
CSGS model relies heavily on the accuracy of the mean or second-moment source data, 
which becomes impractical for flows without reliable information on these data. 

The purpose of this paper is to evaluate the 
performance of recently proposed SGS models and to gain insights into the SGS modeling 
problems by performing wall-resolved LES simulations of channel flows at a
moderate Reynolds number. 
Four SGS models are chosen for this comparative study:
one-k-equation model (kEqn), WALE, sigma, and CSGS. 
The details of these four SGS models are given in Section~\ref{sec:sgs}. 
The simulation parameters and configurations are described in 
Section~\ref{sec:sim}. The results and their discussion are 
presented in Sections~\ref{sec:results} and~\ref{sec:conclusion}. 

\section{SGS models} \label{sec:sgs}
\subsection{kEqn model} \label{subsec:keqn}
The kEqn model~\cite{Yoshizawa_kEq1985,deVille_phdthesis2006} consists of solving a transport 
equation of the SGS kinetic energy
\begin{equation}
  \partial_t k_{SGS} + \textbf{u}\nabla k_{SGS} = 
  (\nu_{SGS}+\nu)\Delta k_{SGS} + 2\nu_{SGS}|S|^2 - \frac{C_{\epsilon}k_{SGS}^{3/2}}{\Delta},
  \label{eq:kSGSEq}
\end{equation}
where $C_k$, $C_{\epsilon}$ are constants. Typically, $C_k = 0.094$ and $C_{\epsilon}=1.048$. 
The terms in the right hand side denote the viscous and turbulent diffusion, 
the gradient diffusion 
(the energy transfer between the filtered and sub-grid scales), and the dissipation, 
respectively. The square root of $k_{SGS}$ is taken as the velocity scale in the 
definition of the SGS viscosity,
\begin{equation}
  \nu_{SGS} = C_k\sqrt{k_{SGS}}\Delta.
  \label{eq:oneEqEddy}
\end{equation}

Compared to the Smagorinsky model~\cite{Smagorinsky_MonWeatherRev1963}, 
the kEqn model takes into account history and 
spatial effects. However, in practice, both models yield very close results. 
In the equilibrium state they are statistically equivalent.

\subsection{WALE model} \label{subsec:wale}
The WALE model was developed by Nicoud and Ducros~\cite{NicoudDucros_1999} 
in order to remedy the imperfection of the Smagorinsky model. In the Smagorinsky 
model, the SGS viscosity does not go to zero as the wall is approached. This 
causes an overestimation of the SGS dissipation and the log-layer mismatch problem. 
The WALE model is constructed from the invariants 
of the square of the velocity gradient tensor and achieves the correct scaling 
behavior near the wall. The WALE model preserves the property of locality, meaning 
that only local quantities are required to evaluate the model at any point 
in space and time. The model reads 
\begin{subequations}
  \begin{align}
    \nu_{SGS} &= (C_w\Delta)^2\mathcal{D}_w(\textbf{u}), \\
    \mathcal{D}_w &= \frac{(S_{ij}^dS_{ij}^d)^{3/2}}{(S_{ij}S_{ij})^{5/2}+(S_{ij}^dS_{ij}^d)^{5/4}},
  \end{align}
\end{subequations}
where $\mathcal{D}_w$ is a differential operator, $S_{ij}^d$ is the traceless symmetric 
part of the square of the velocity gradient tensor, and $C_w \simeq 0.165$. 

\subsection{Sigma model} \label{subsec:sigma}
Proposed also by Nicoud et al.~\cite{NicoudLee_pof2011}, 
the Sigma model is an advanced variant of WALE, embracing 
a broader range of applicability. Instead of being based on the invariants of 
the velocity gradient functionals, the Sigma model employs the singular values 
of the velocity gradient tensor $\textbf{g} = u_{i,j}$. The Sigma model is
\begin{subequations}
  \begin{align}
    \nu_{SGS} &= (C_{\sigma}\Delta)^2\mathcal{D}_{\sigma}(\textbf{u}), \\
    \mathcal{D}_{\sigma} &= \frac{\sigma_3(\sigma_1-\sigma_2)(\sigma_2-\sigma_3)}{\sigma_1^2},
  \end{align}
\end{subequations}
where $\sigma_{1,2,3}$ denote the three singular values of $\textbf{g}$ and 
satisfy $\sigma_1 > \sigma_2 > \sigma_3 > 0$. By definition, the sigular values of $\textbf{g}$ 
are the square roots of the eigenvalues of $\textbf{g}^t\textbf{g}$. The model constant 
$C_{\sigma}$ is typically taken to be 1.35. Note that Sigma model is designed 
for boundary layer flows with smooth walls. With rough walls, the near wall scaling 
may be different from that of the smooth wall and approriate adjustment of the model is 
required to better capture the near-wall dynamics. 

\subsection{CSGS model} \label{subsec:cles}
Originally used in optimizaton problems, constraints were recently introduced 
to model the SGS flow motions~\cite{ShiChen_pof2008}. 
Given that constraints reflect basically the 
current available knowledge of a problem, the philosophy of CSGS is to use a priori
known flow statistics to guide the behavior of a model. 
This methodology was proposed for LES simulations~\cite{ShiChen_pof2008,ChenShi_jfm2012} 
and has already been applied in many situations~\cite{JiangChen_pof2014,ZhaoChen_pof2014}. Constraints can 
be applied to different physical quantities, such as Reynolds shear stress or SGS dissipation, 
depending on the currently available knowledge and the importance of the quantities. 
Following the proposal in~\cite{ChenShi_jfm2012}, in this paper a constraint is placed 
on the Reynolds shear stress, since its inaccurate prediction 
is thought to be responsible for the log-layer mismatch near the wall. 
The constraint can be obtained from high-fidelity DNS, 
experimental/theoretical results, or even a RANS model in certain circumstances. 

The idea of the Reynolds-stress constrained SGS model can be expressed by the following 
decomposition of the SGS stresses, 
\begin{equation}
  \tau_{ij} = \underbrace{R_{ij} - R_{ij}^{LES}}_{\langle \tau_{ij} \rangle} + \tau_{ij}',
\end{equation}
where $R_{ij}$ is the actual knowledge of the Reynolds stress of the physical velocity field, 
$R_{ij}^{LES}$ is the Reynolds stress of the resolved velocity field in LES, and 
$\tau_{ij}$ is the sub-grid scale stress. Since $R_{ij}$ is known and $R_{ij}^{LES}$  
can be evaluated by $\langle u_i'u_j'\rangle$ ($\langle\cdot\rangle$ denotes 
the ensemble average operator), the only term to be modeled is the 
fluctuating part $\tau_{ij}'$. 

There are many different implementations of $\tau_{ij}'$. In\added{ the paper by Chen et al.}~\cite{ChenShi_jfm2012}, 
$\tau_{ij}'$ is evaluated by the dynamic Smagorinky procedure. 
In this paper, a slightly different approach is employed: 
choose a baseline model (e.g., kEqn), evaluate the SGS stress $\tau^{kEqn}$ of the baseline model, 
and calculate an additional term $(R_{ij} - R_{ij}^{LES}) - \langle \tau_{ij}^{kEqn}\rangle$, 
\added{namely, $\tau_{ij} = \tau_{ij}^{kEqn} + [(R_{ij} - R_{ij}^{LES}) - \langle \tau_{ij}^{kEqn}\rangle]$}. 
The additional term is the difference of the two mean SGS stresses, one from the existing 
knowledge and another from the baseline model. Actually, $\tau_{ij}'$ is modeled by 
the fluctuating part of the baseline SGS stress, $\tau' = \tau^{kEqn} - \langle \tau^{kEqn}\rangle$, 
which fullfils the necessary condition $\langle \tau'\rangle = 0$. \added{The derivation is 
as following: $\tau' = \tau - \langle \tau \rangle = (R_{ij} - R_{ij}^{LES}) - \langle \tau_{ij}^{kEqn}\rangle + \tau_{ij}^{kEqn} - [(R_{ij} - R_{ij}^{LES})] = \tau_{ij}^{kEqn} - \langle \tau_{ij}^{kEqn}\rangle$.}
This implementation is non-intrusive, meaning that it does not require any modification of 
the pre-existing LES code. For modular programming language, only a subroutine or module 
is needed, which calculates the additional term.

\section{Simulation details} \label{sec:sim}
The computational domain size of the channel flow is taken to 
be $(2\pi \times \pi \times 2)H$, where 
$H$ is the half-channel height. The coordinates (X,Y,Z) denote the streamwise, the spanwise 
and the wall normal direction, respectively. The spatial resolution 
is $(96 \times 96 \times 96)$. The cell size is uniform in the X and Y directions, and it is 
linearly 
decreasing in the Z direction towards the wall. The difference between the maximum and 
minimun cell size in the Z direction is 
$\Delta z_{max}/\Delta z_{min} \simeq 10$. The wall-unit size $z^+ =zu_*/\nu$ of the first point 
is about 1. \added{The wall-unit mesh size in X and Y direction are about 25.9 and 12.9, respectively. }
To compare with the DNS results in~\cite{Moser_pof1999}, 
the wall Reynolds number is $Re_\tau = u_* H/\nu = 395$, which is imposed by 
a constant external 
pressure gradient $dP/dx = u_*^2 = 6.24\times 10^{-5}$, 
assuming $\nu=2\times 10^{-5} \text{ m}^2/\text{s}$ and $H=1$ m. 
The no-slip boundary condition is applied at the top and bottom walls, 
while periodic boundary conditions are used 
at the other boundaries. Initial conditions for the velocity field consist of 
streak-like structures near the wall to reduce the initial relaxation time. 
For the constrained SGS model, the Reynolds-stress 
constraints are obtained from the DNS results~\cite{Moser_pof1999} 
and are imposed in the near-wall region ($z^+ < 40$ or $z < 0.1H$), 
called hereafter the constrained region. 
The baseline model for the constrained SGS model is the kEqn model. 
In the unconstrained region, the baseline model is used as the SGS model. 
As pointed out in~\cite{ChenShi_jfm2012}, the location of the interface 
between the constrained and unconstrained region as well as the baseline model 
do not influence significantly the results. \removed{In~\ref{app:sens}, t}The sensitivity 
of the results to the choice of baseline model and the constrained region 
is \changed{tested}{discussed} \added{in the next section}. \removed{
The results are consistent with~\cite{ChenShi_jfm2012}. }

Simulations were performed with OpenFOAM 3.0.1~\cite{openfoam_2015}. The CSGS and Sigma model 
were implemented with this version. \verb+PimpleFoam+ is used to integrate numerically the filtered 
Navier-Stokes equations. The 2nd-order backward difference is employed for the time discretization and 
the 2nd-order Gauss-type schemes are used for the spatial discretization. 
The details of the numerical methods in OpenFOAM and their validation are presented 
in~\cite{RobertsonWalters_caf2015}. 

\section{Results} \label{sec:results}
Flow statistics, including the SGS viscosity and stress, the mean velocity, 
the second moments and the fluctuating vorticities, are analyzed 
for each SGS models and are compared with the DNS results by Moser et al.~\cite{Moser_pof1999}.
All quantities are averaged in time and in the wall-parallel directions 
(for simplicity the averaging operator is omitted). \added{The average is 
performed after reaching the statistically stationary state for a duration of 10000 s, 
about 250 turnovers using the central-line velocity.} 
The vertical (Z) profiles along half channel height are shown in the figures 
(Figs. 1-6, 8) of this section, where 
the lines with symbols denote the SGS models and the dashed ones are the DNS results.
The shaded region in the plots corresponds to the buffer layer, in which the distance from the 
wall is in the range $z^+ \in [5, 40]$ or $z \in [0.013,0.1]H$. 
Below the buffer layer is the viscous sublayer while above it are the logarithmic 
layer and the outer layer. The buffer layer is shaded, since, as will 
be shown in this section, the SGS models play a key role in this region.

\begin{figure}[!ht]
  \centering
  \includegraphics[width=0.45\textwidth]{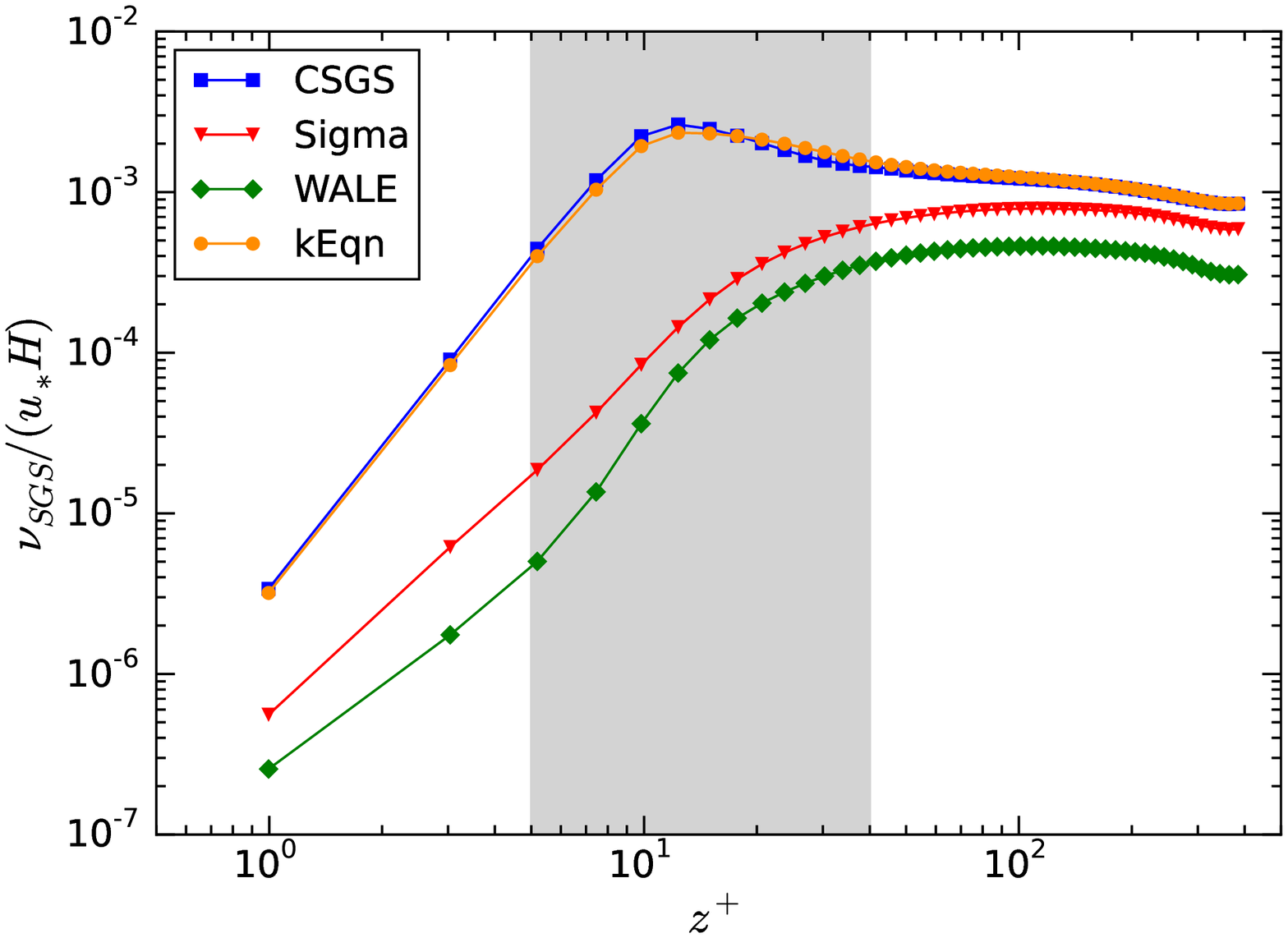}
  \includegraphics[width=0.45\textwidth]{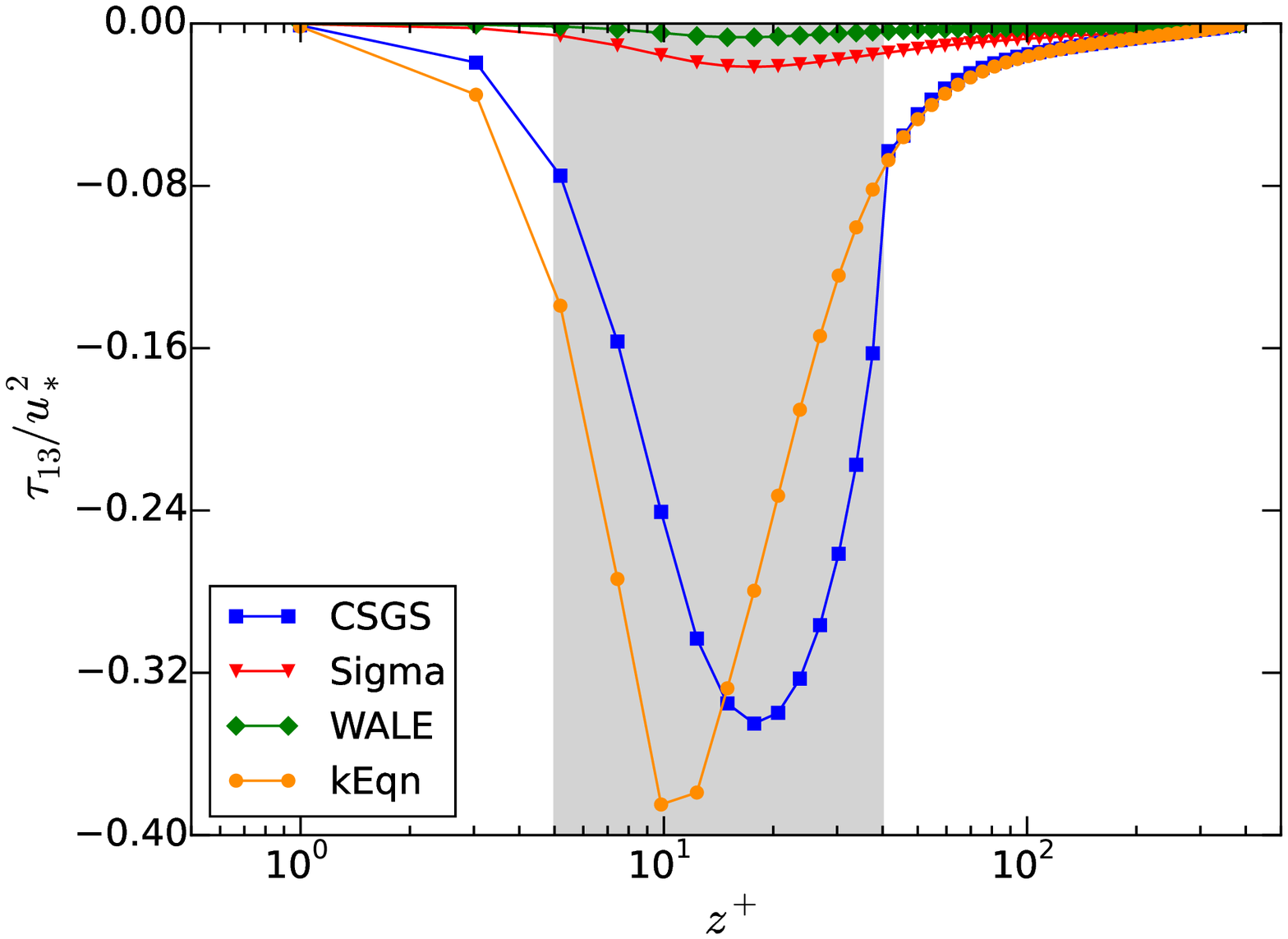}
  \caption{Profiles of the normalized SGS viscosity $\nu_{SGS}$ (left) and stress $\tau_{13}$ 
    (right). The color markers correspond to different SGS models. 
    The shaded region is the buffer layer ($z^+ \in [5,40]$ or $z \in [0.013,0.1]H$). 
    The colors and the shaded region represent the same variables in other figures.}
  \label{fig:nuSGS}
\end{figure}

The SGS viscosities and stresses are the direct results of the SGS models, which are 
hence firstly studied. 
Figure~\ref{fig:nuSGS} shows the profiles of the SGS viscosity $\nu_{SGS}$ and the 
major component $\tau_{13}$ of the SGS stresses (other components are negligibly small). 
\added{The $\nu_{SGS}$ in the CSGS model denotes the $\nu_{SGS}$ in the baseline kEqn model.}
In the viscous sublayer, 
the SGS viscosity scales in all models with the distance as $\nu_{SGS}\sim z^3$, which agrees 
with the theoretical expectation~\cite{Pope_turbFlow2000}. Nevertheless, the magnitudes 
differ by up to one order of magnitude, 
the smallest being from the WALE model and the biggest from the kEqn and CSGS model. 
Given that all models considered here are based on the eddy-viscosity assumption, 
$\tau = -2\nu_{SGS}|S|$, the difference in $\nu_{SGS}$ is translated into the SGS stress, 
as clearly shown in Fig.~\ref{fig:nuSGS} (right). 
Interestingly, in the buffer layer, the differences of the SGS 
viscosity and stress among models are significantly higher than in other regions. 
Moreover, since the baseline model of the CSGS model is kEqn and the CSGS can be viewed simply 
as a correction to the kEqn model, the closeness in $\nu_{SGS}$ between these two models 
is expected and indicates that the constraint has a negligible effect on the SGS kinetic 
energy and viscosity. However, the constraint has a large impact on the SGS stress in the 
constrained region ($z^+ < 40$), as is shown in the profile of $\tau_{13}$. 
How the turbulence statistics are affected by the difference in the 
magnitude of $\nu_{SGS}$ and the correction from the constraint will be explained in the 
remainder of this section.

\begin{figure}[!ht]
  \centering
  \includegraphics[width=0.7\textwidth]{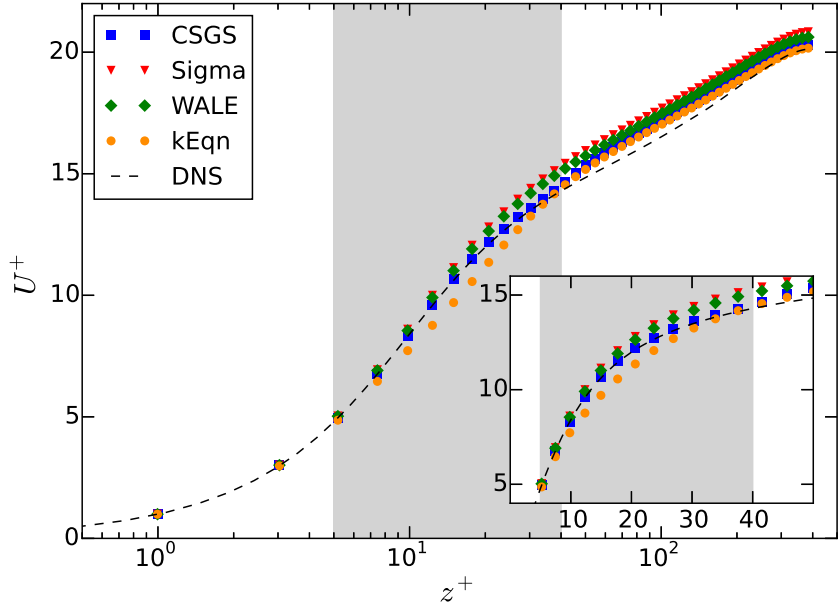}
  \caption{Profiles of the mean streamwise velocity in wall units. 
    The dashed line is the DNS data from 
    ~\cite{Moser_pof1999}. Inset is a replicate 
    of the buffer layer with X-axis in linear scale; 
    note that the mean velocity based on the CSGS model best matches the DNS result.}
  \label{fig:U}
\end{figure}

The profiles of the mean streamwise velocity in wall units, $U^+ = U/u_*$, 
are shown in Fig.~\ref{fig:U}. 
All SGS models achieve a mean-velocity profile that agrees within a relative error 
of about 6\% with the the DNS results. Moreover, a remarkably close match is found in 
the viscous sublayer for all models. 
In the logarithmic layer and outer layer, all profiles display an overestimation of 
the mean velocity, i.e., the log-layer mismatch problem. 
The mismatch in Sigma and WALE is negligibly greater than those 
in kEqn and CSGS. However, in the buffer layer, different models behave in distinct ways. 
The kEqn model underestimates the mean velocity while Sigma and WALE gradually overpredict it. 
The CSGS achieves the best mean velocity profile. In the whole constrained region, 
the mean velocity profile is in excellent agreement with the DNS results, 
since the constraint applied in the CSGS simulation is the mean Reynolds 
stress from the DNS simulation. 

\begin{figure}[!ht]
  \centering
  \includegraphics[width=0.45\textwidth]{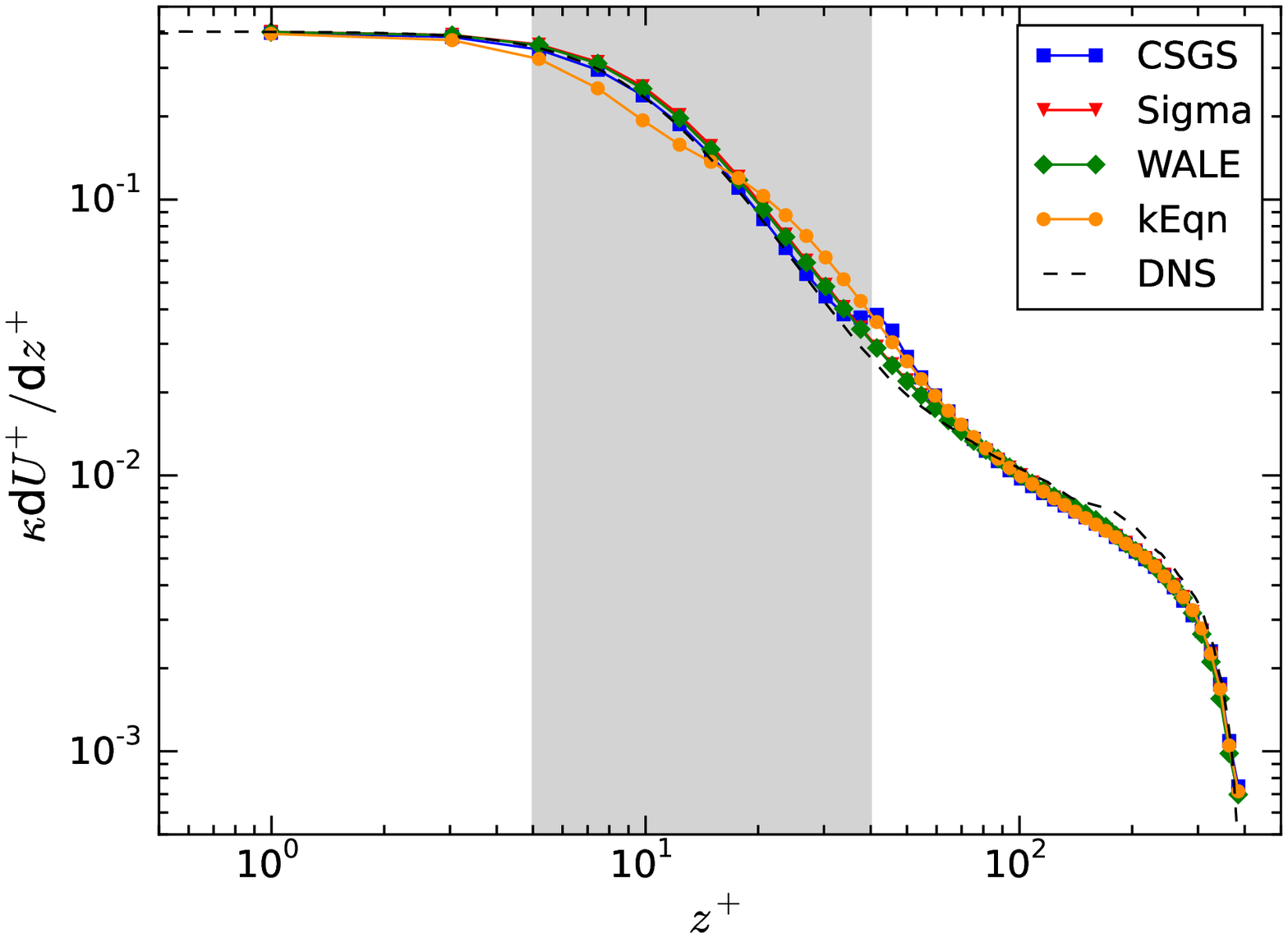}
  \includegraphics[width=0.45\textwidth]{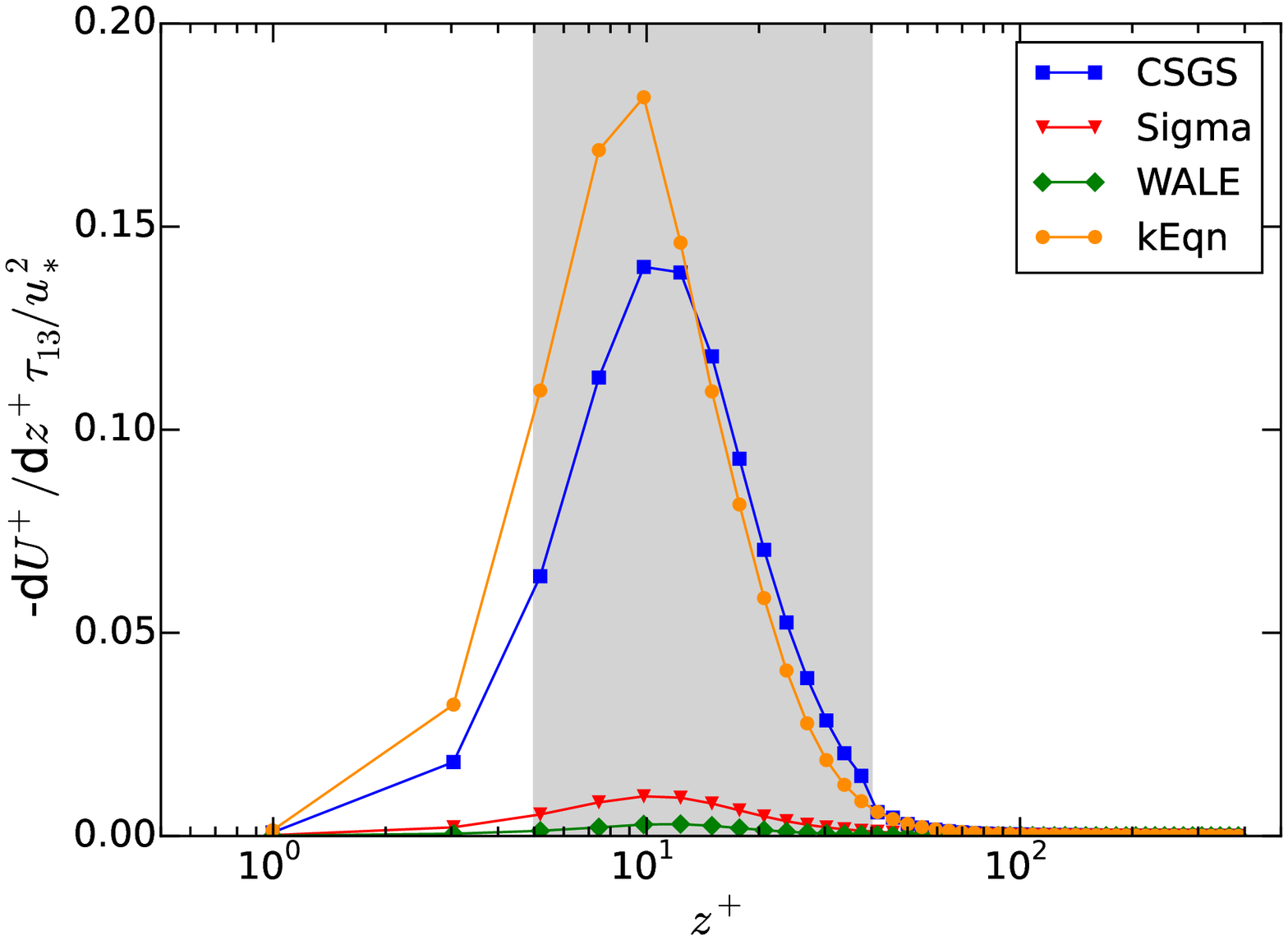}
  \caption{Profiles of the normalized velocity gradient (left) and the normalized mean-flow SGS 
    dissipation $\tau_{13}\text{d}U/\text{d}z$ (right).}
  \label{fig:sgsDissip}
\end{figure}

A more dynamically relevant quantity is the mean velocity gradient, which is plotted in 
Fig.~\ref{fig:sgsDissip} (left). Except for the kEqn model in the buffer layer and 
the CSGS model near the top of the constrained region, all profiles agree very well 
with the DNS results. The mismatch of the mean velocity profiles in the log layer and 
outer layer is also significantly reduced in the velocity gradient. 
It seems that the root of the log-layer mismatch in the mean velocity profile in the simulations
lies in the inferior model performance in the buffer layer. Moreover, unlike 
in previous studies where the mismatch problem is claimed to be induced by the inaccurate SGS 
dissipations~\cite{PorteAgel_jfm2000, KawaiLarsson_pof2012, WuMeyers_pof2013}, 
the profiles of the mean-flow SGS dissipations here (Fig.~\ref{fig:sgsDissip} right) 
does not show a noticeable correlation with the observed mismatch: although the SGS 
dissipation in both kEqn and CSGS are very high compared to others, the predictions of the 
mean velocity and its gradient are very close. 

Consider now the force balance in the mean flow. Averaging the momentum equation 
for the streamwise velocity in time and in the horizontal (X and Y) directions, one obtains

\begin{equation}
  \centering
  uw + \tau_{13} - \nu\frac{\text{d}U}{\text{d}z} = u_*^2(\frac{z}{H} -1),
  \label{eq:uEqn}
\end{equation}
where the \added{mean} SGS stress \changed{can be written}{is defined} as \changed{$\tau_{13} = -\nu_{SGS}\text{d}U/\text{d}z$}
{$\tau_{13} = -\langle \nu_{SGS}\text{d}u/\text{d}z \rangle$}. 
\added{Note that the averaging operator $\langle \cdot \rangle$ is omitted in above equation. }
The derivation of this equation assumes temporal and horizontal statistical homogeneity, 
which is satisfied in the statistically stationary channel flow. 
\added{Note that this relation is an exact integral form of the momentum equations and 
shows that, indepentent from the SGS modeling, the total stress is a linear function of the distance 
to the wall. Any deviation from this behavior is not due to the SGS model but to other numerical issues.}

\begin{figure}[!ht]
  \centering
  \includegraphics[width=0.7\textwidth]{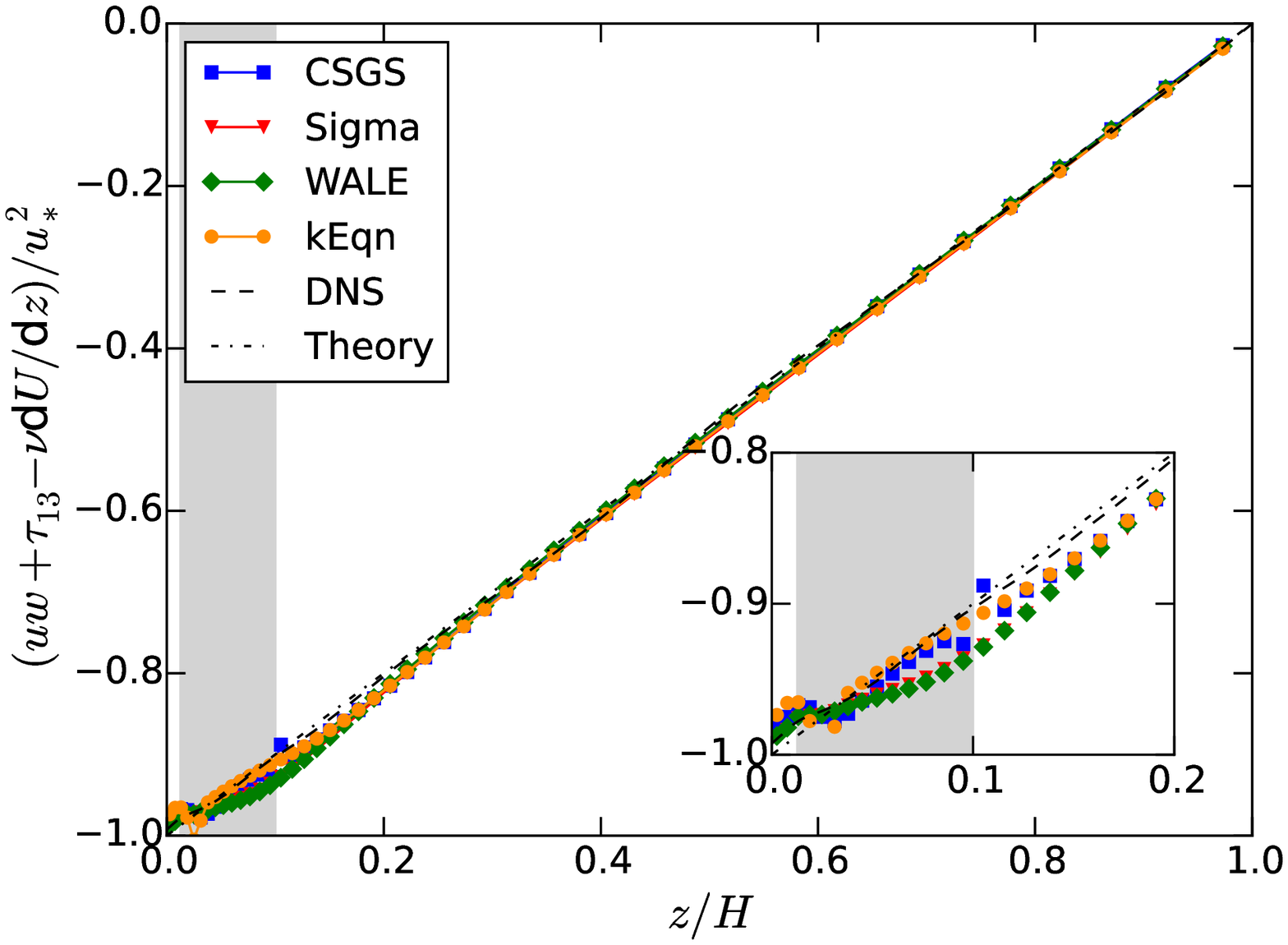}\\
  \includegraphics[width=0.45\textwidth]{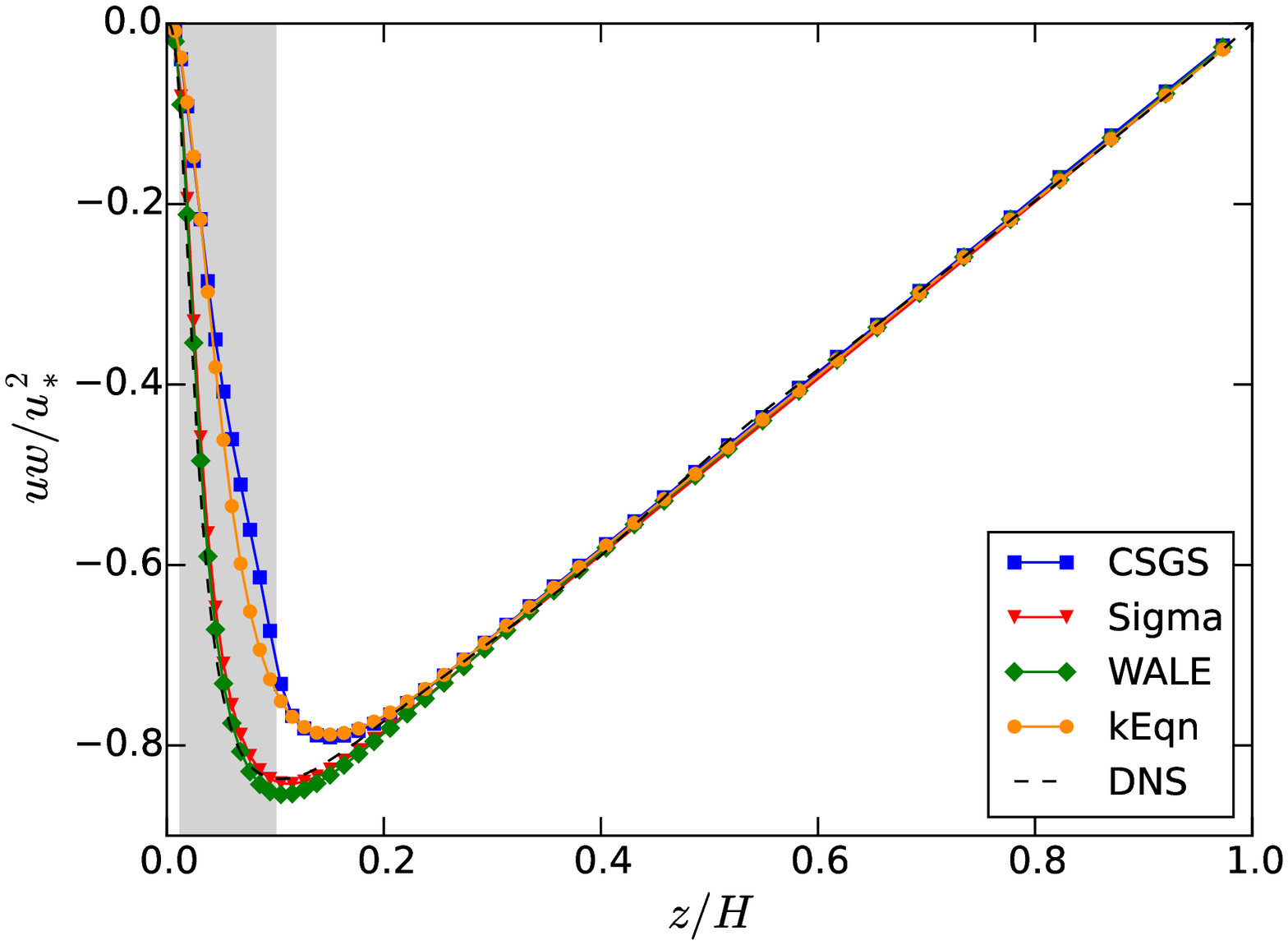}
  \includegraphics[width=0.45\textwidth]{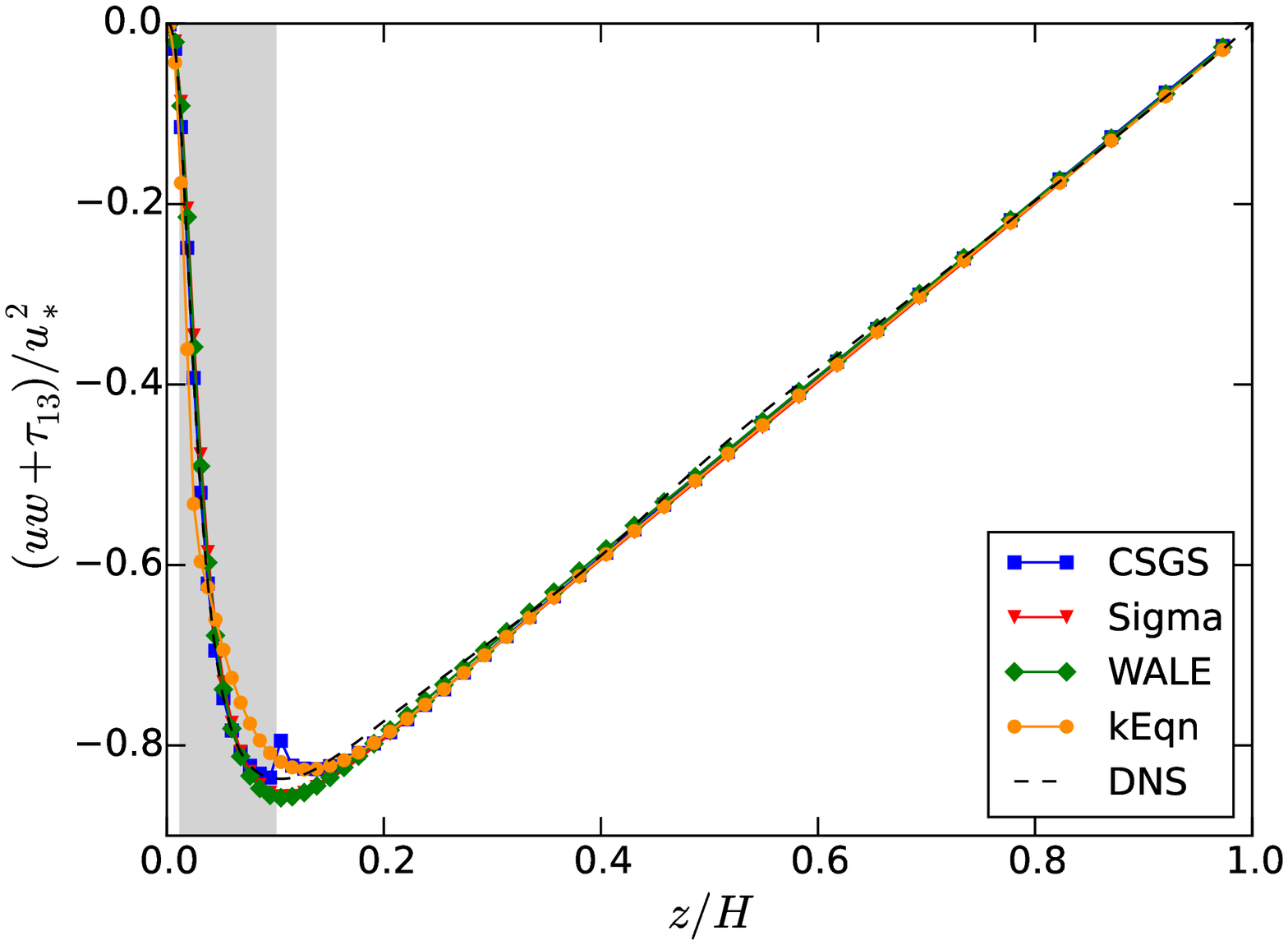}
  \caption{Profiles of the total shear stress (top), the resolved (bottom left) and 
    full (bottom right) Reynolds shear stresses. The dash-dotted line corresponds to 
    the theoretical result Eq.~\eqref{eq:uEqn}. The inset in the top figure is a zoom-in of 
    the near-wall region.}
  \label{fig:stress}
\end{figure}

The total shear stress, and
the resolved and full Reynolds shear stresses, are plotted in Fig.~\ref{fig:stress}. 
Here only the $xz$ or $13$ component is shown. The other two shear components 
are theoretically zero. All \added{total shear} profiles agree very well with theoretical 
results, except in the near-wall region $z<0.2H$. In this region, the total shear profiles in
the WALE and Sigma model deviate slightly further from the theoretical curve than in CSGS and 
kEqn, overall within 2\% relative error\removed{, including certain unavoidable numerical errors}. 
\added{This deviation due to numerical errors is very small compared to that for other statistics, 
which hints a small numerical viscosity in the simulations.}

In the $uw$ profile, WALE and Sigma 
are much closer to the DNS curve, while in the $uw+\tau_{13}$ profile all models perform very 
closely. From all above results, it seems that, among the variables in Eq.~\eqref{eq:uEqn}, 
$uw$ correlate highly with $\nu_{SGS}$, while the velocity gradient $\text{d}U/\text{d}z$ is a result 
of much more complicated mechanism, including the governing equations and the SGS models. 
This complexity makes it extremely difficult to solve the log-layer mismatch problem, 
and also indicates that ensuring correct SGS and Reynolds shear stresses is only a necessary 
but not a sufficient condition for accurately predicting the mean flow.

\begin{figure}[!ht]
  \centering
  \includegraphics[width=0.45\textwidth]{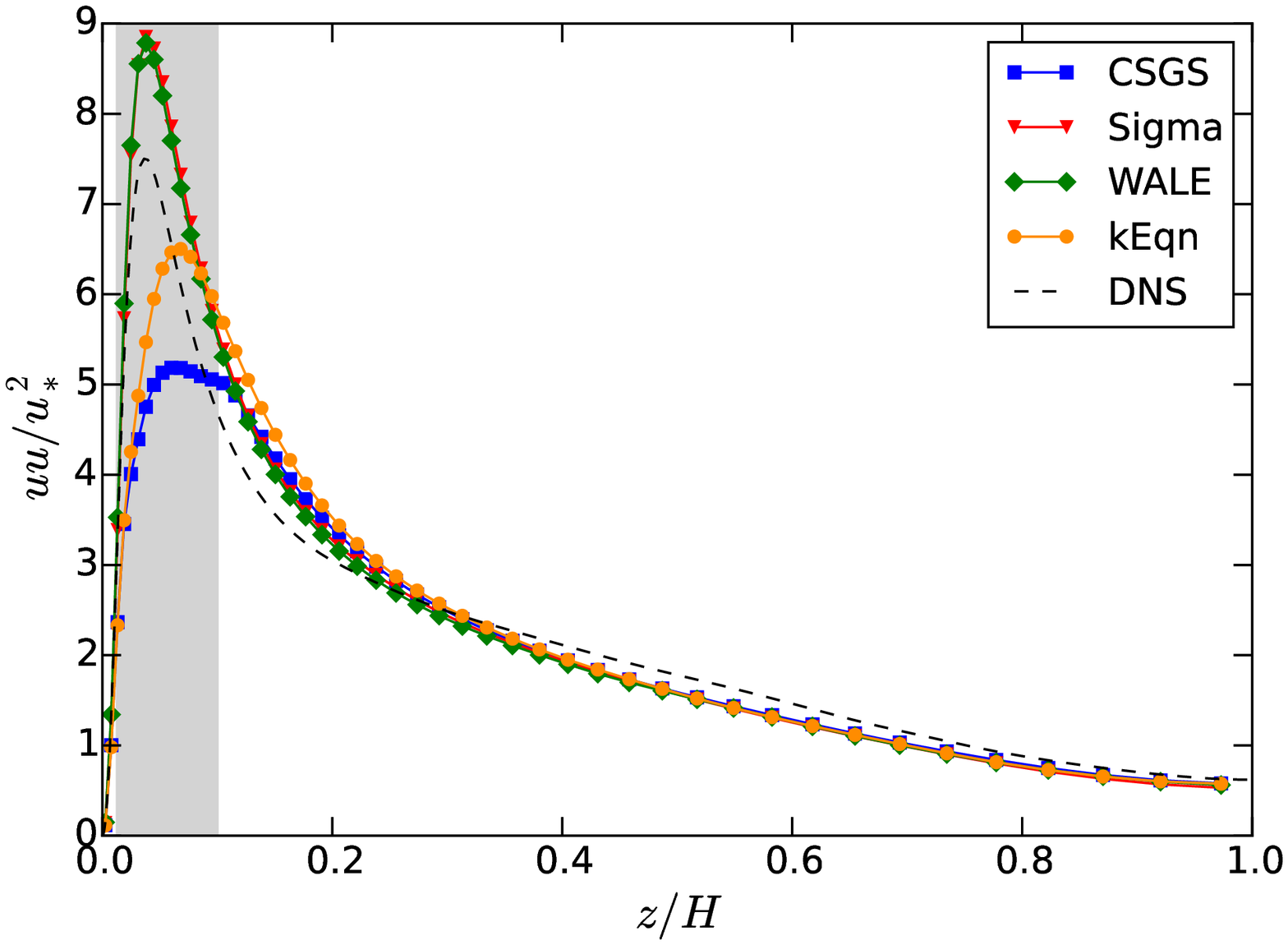}
  \includegraphics[width=0.45\textwidth]{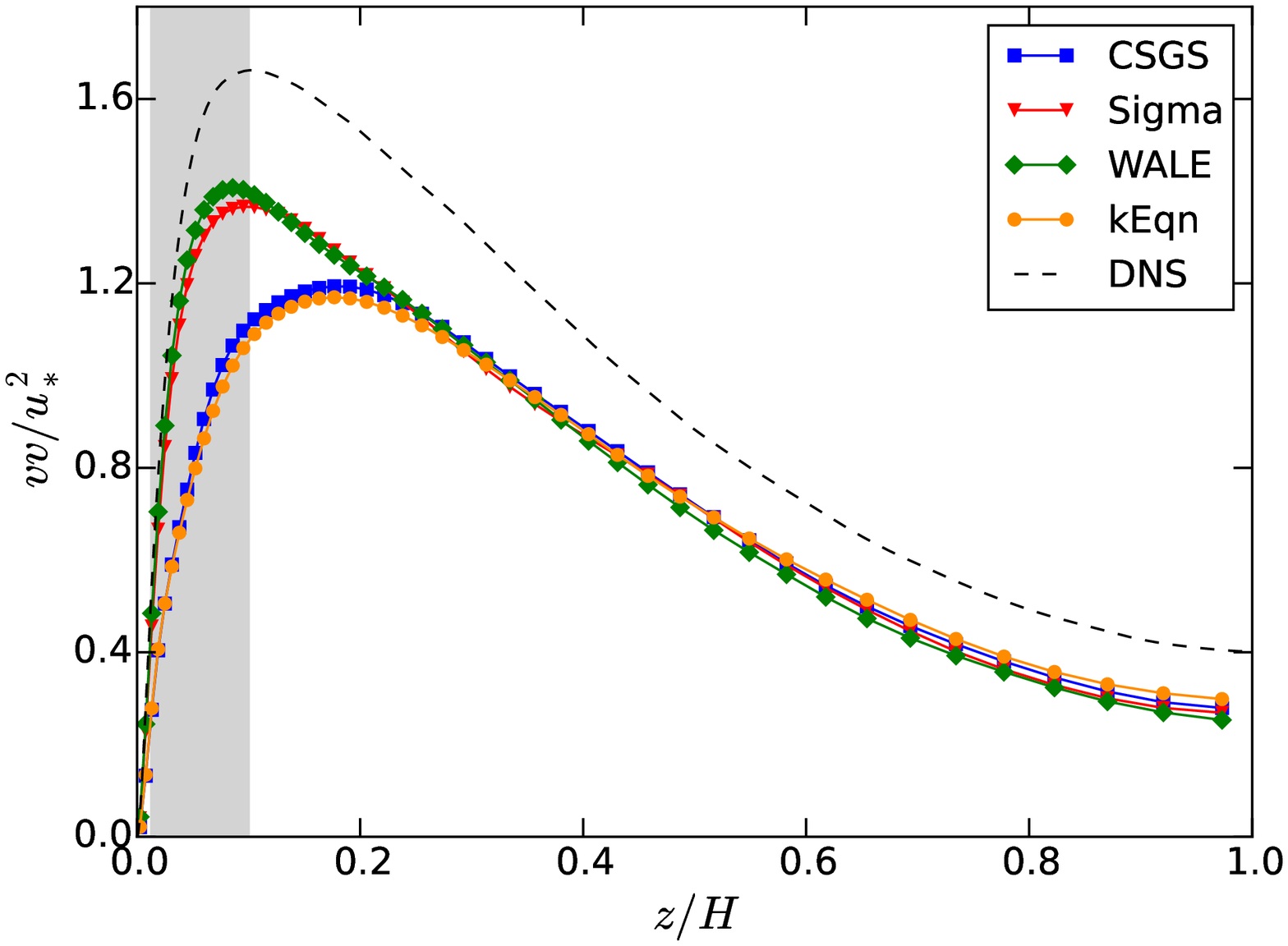}
  \includegraphics[width=0.45\textwidth]{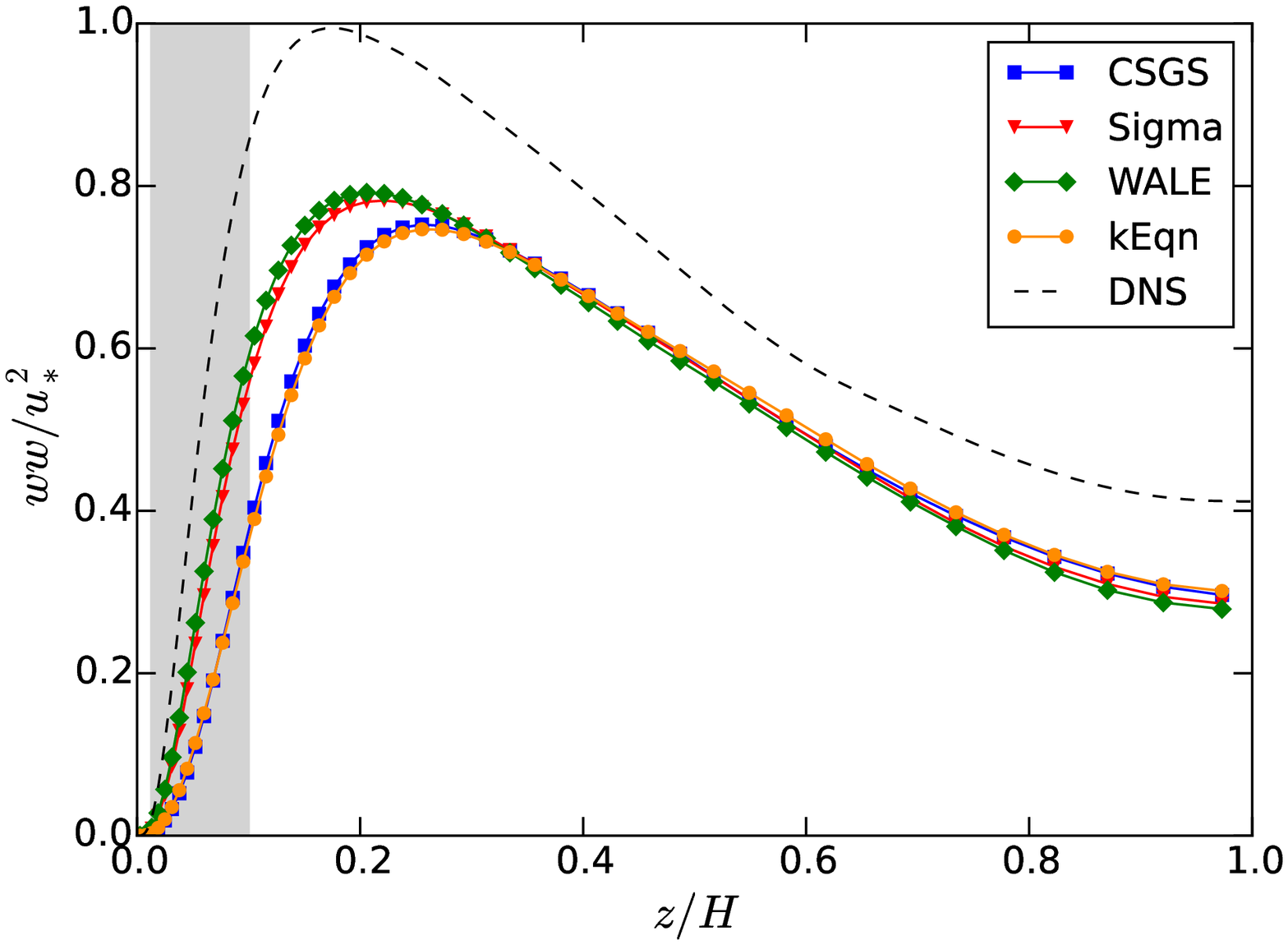}
  \caption{Profiles of the resolved veloicty variances.}
  \label{fig:uu}
\end{figure}

The turbulence intensities are characterized by the resolved velocity variances, 
which are plotted in Fig.~\ref{fig:uu}. Except for the streamwise variance $uu$ in 
WALE and Sigma, all other profiles show an underestimation compared to the DNS level and 
are almost the same in the outer layer. 
In the near wall region, WALE and Sigma overestimate $uu$ by up to 20\% and 
resolve the variances 
$vv$ and $ww$ better than the other models. It is interesting that the CSGS model performs 
even worse in $uu$ than its baseline model kEqn, 
although CSGS predicts a better mean velocity profile\changed{, as shown in Fig.~\ref{fig:U}}{(see Fig.~\ref{fig:U})}. 
\added{Note that the inaccurate predictions of variances are not necessarily due to the incorrectly modeled SGS dissipation, 
which can also be induced by the nonnegligible numerical viscosity inherent in the finite volume method.}
\changed{Again,}{Overall, it is shown that} different models result in significant differences in the buffer layer, including  
the location of the peaks. \removed{Moreover, the inaccurate predictions of variances are 
not necessarily due to the incorrectly modeled SGS dissipation: the SGS dissipations in 
WALE and Sigma are very small as shown in Fig.~\ref{fig:sgsDissip}. }

\begin{figure}[!ht]
  \centering
  \includegraphics[width=0.45\textwidth]{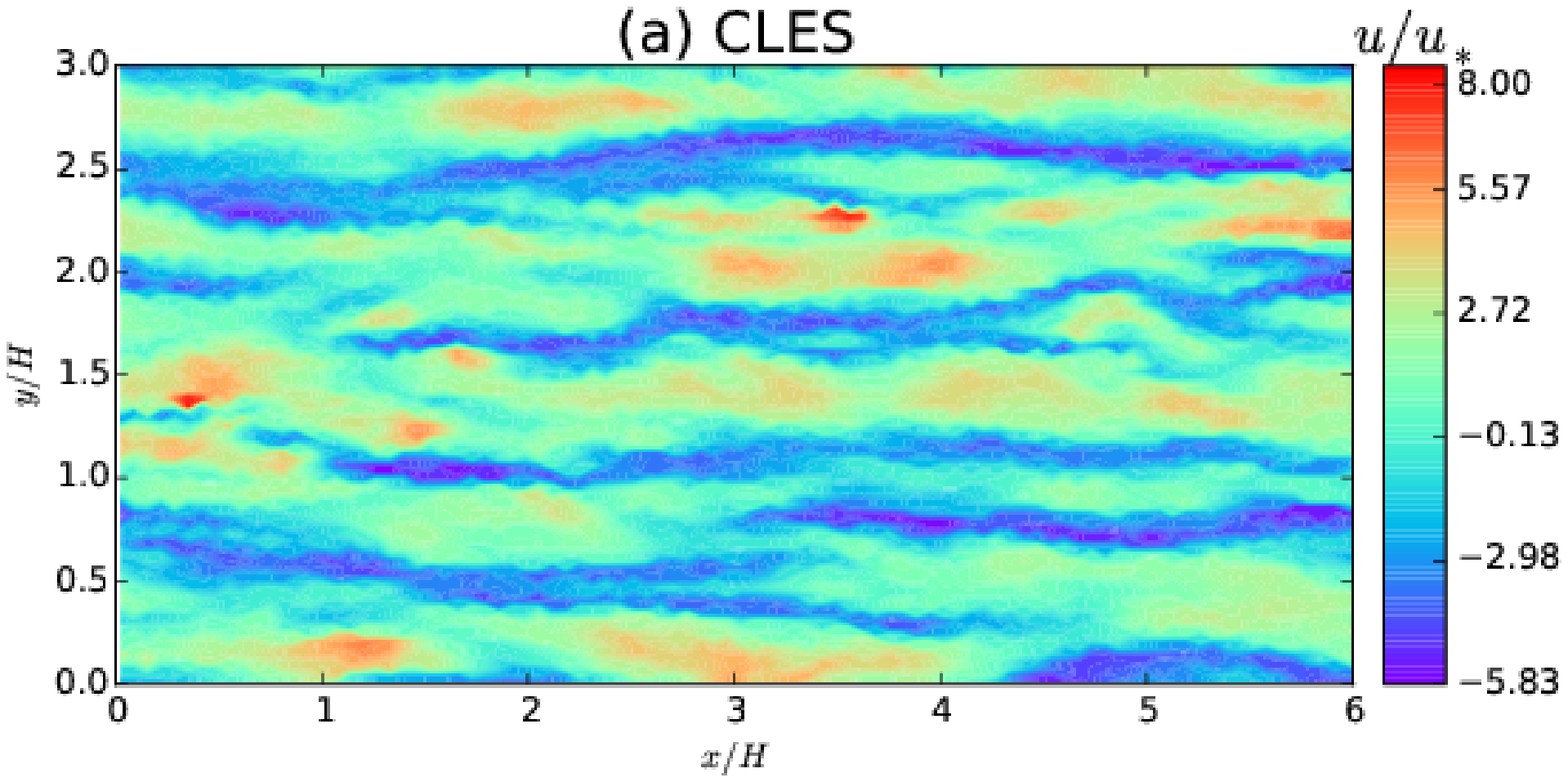}
  \includegraphics[width=0.45\textwidth]{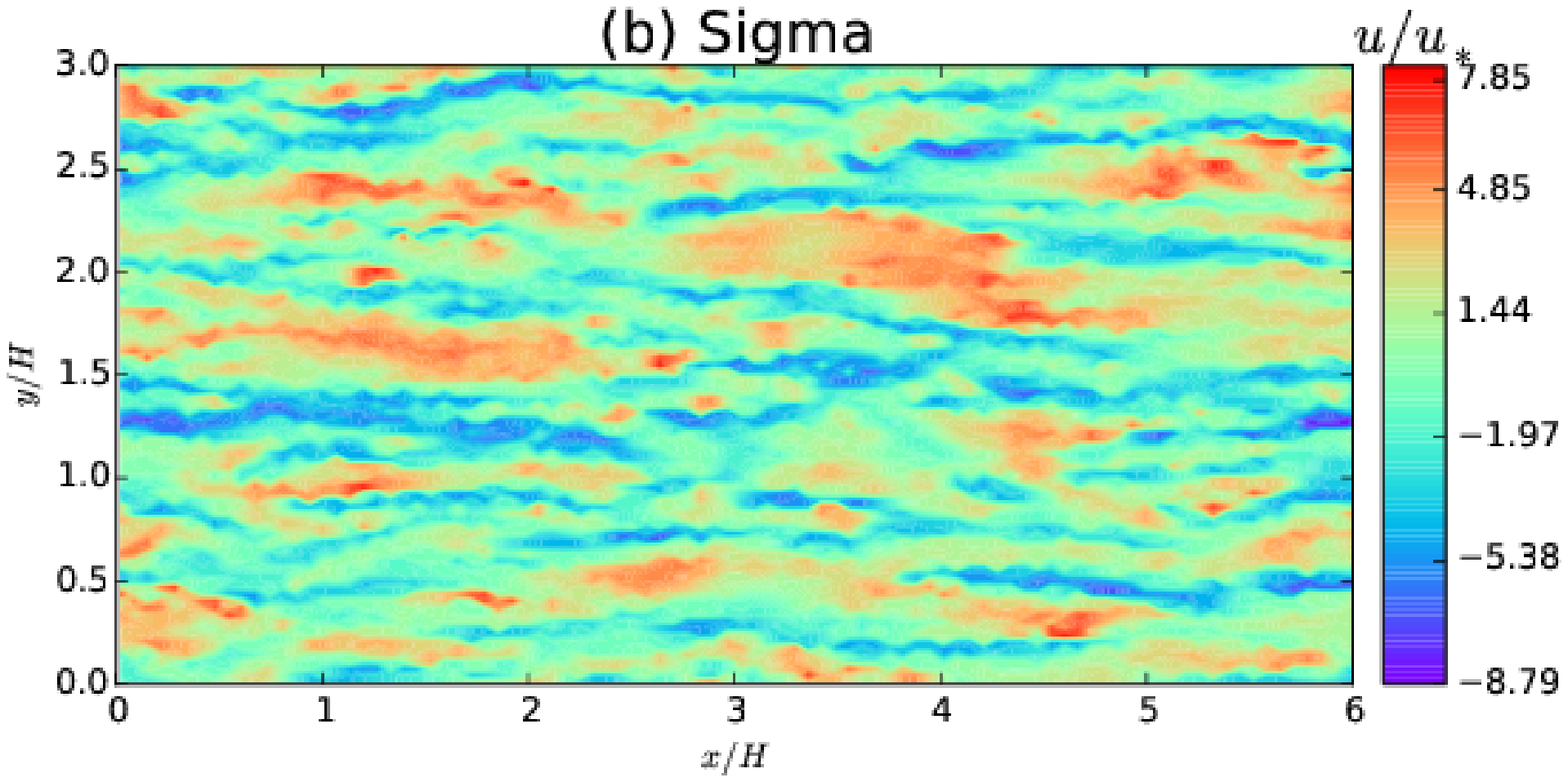}
  \includegraphics[width=0.45\textwidth]{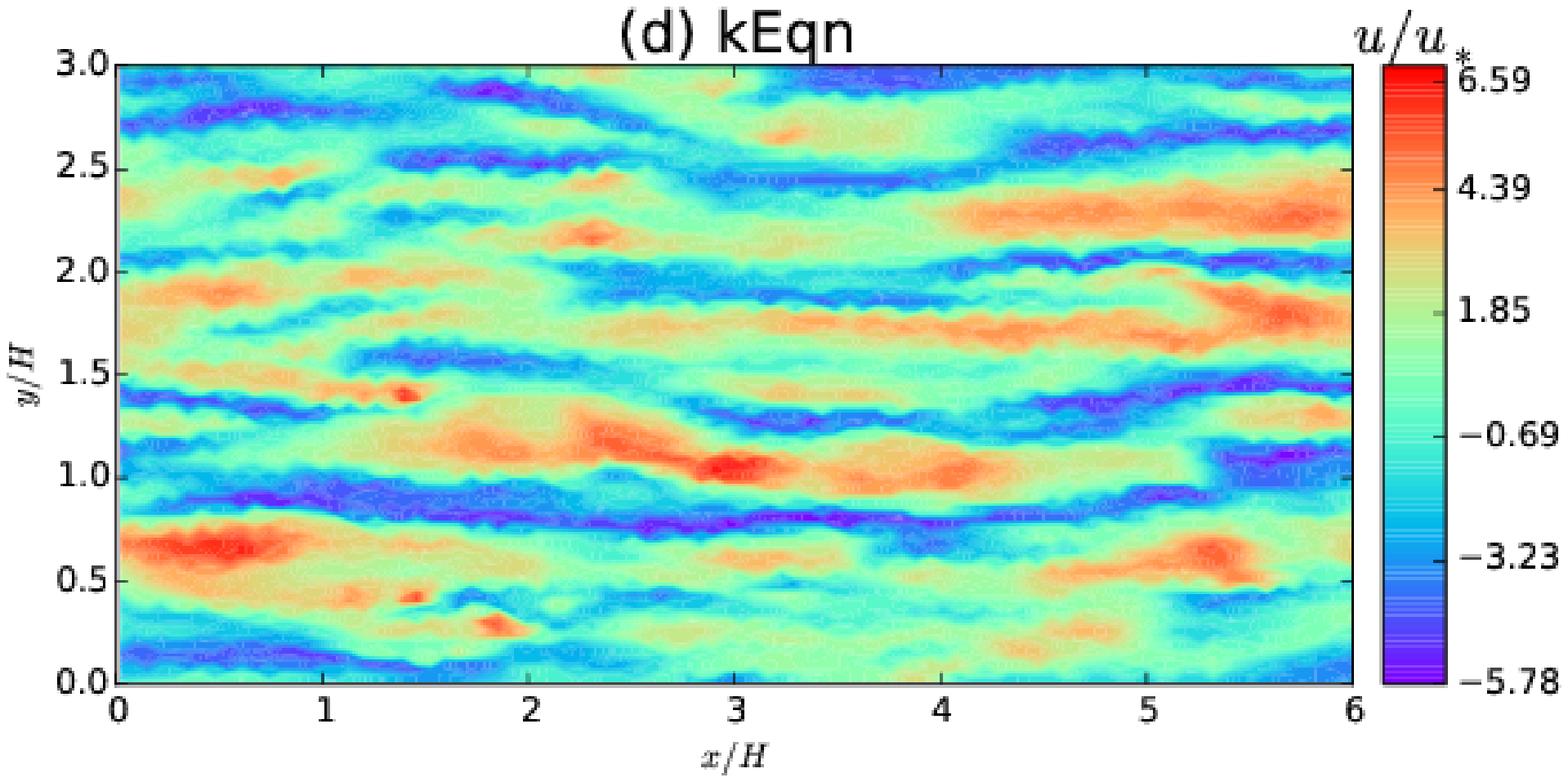}
  \includegraphics[width=0.45\textwidth]{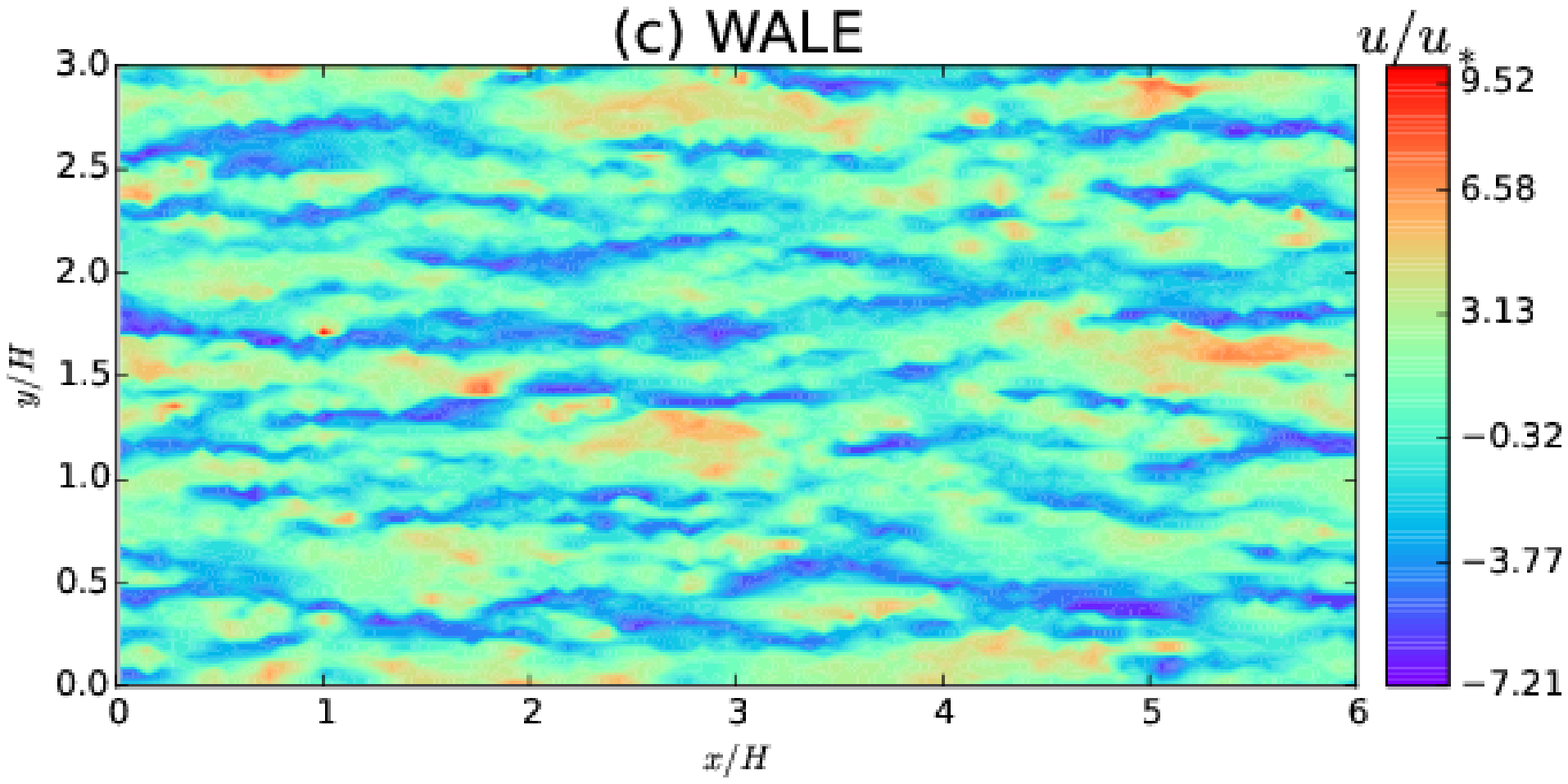}
  \caption{Contour of instantaneous velocity fluctuation at $z/H=0.05$ 
    (in the buffer layer). 
    (a) CLES, (b) Sigma, (c) WALE, (d) kEqn. Turbulent structures are much finer for 
    WALE and Sigma near the wall.}
  \label{fig:contour005}
\end{figure}

\begin{figure}[!ht]
  \centering
  \includegraphics[width=0.45\textwidth]{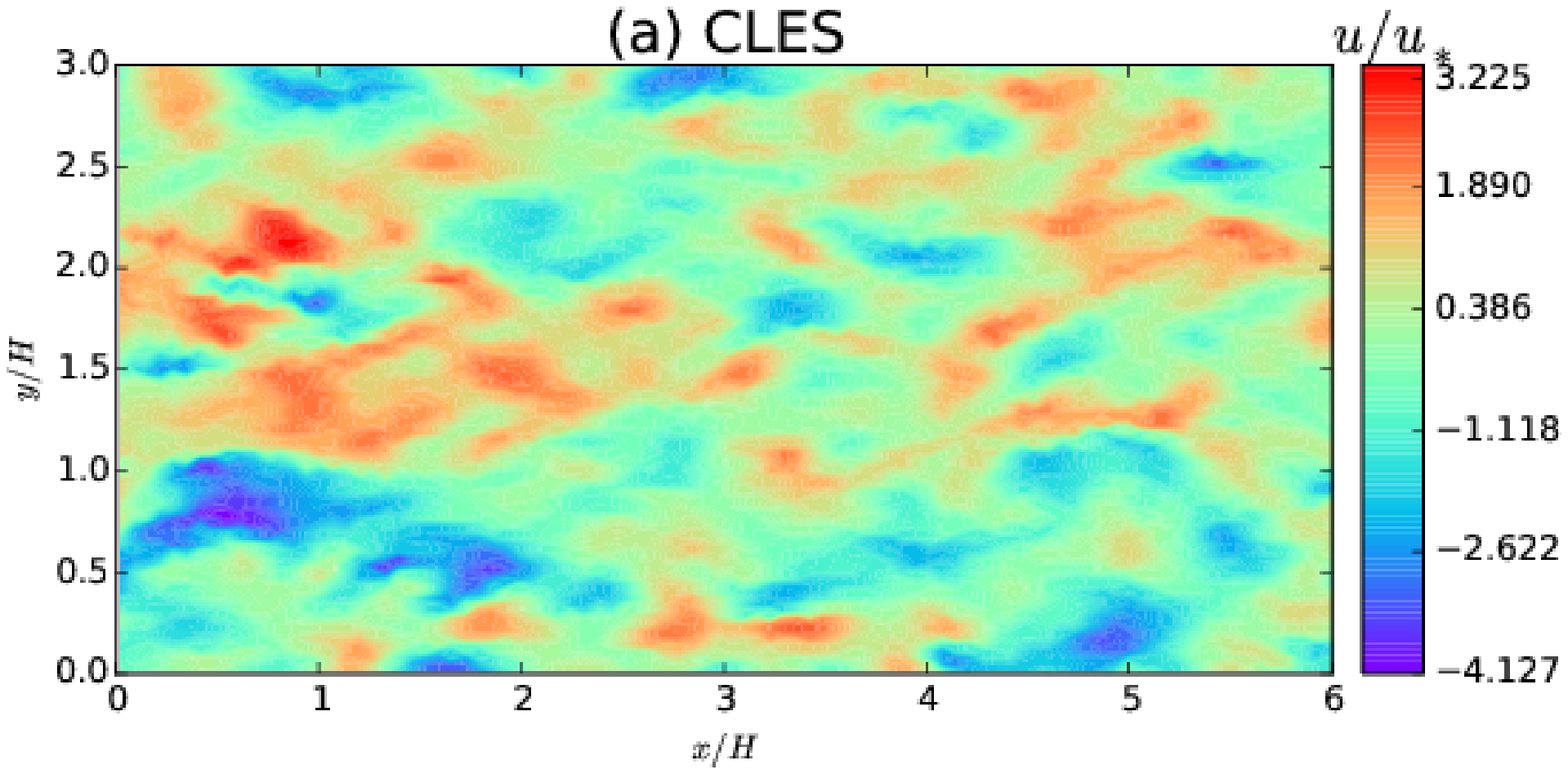}
  \includegraphics[width=0.45\textwidth]{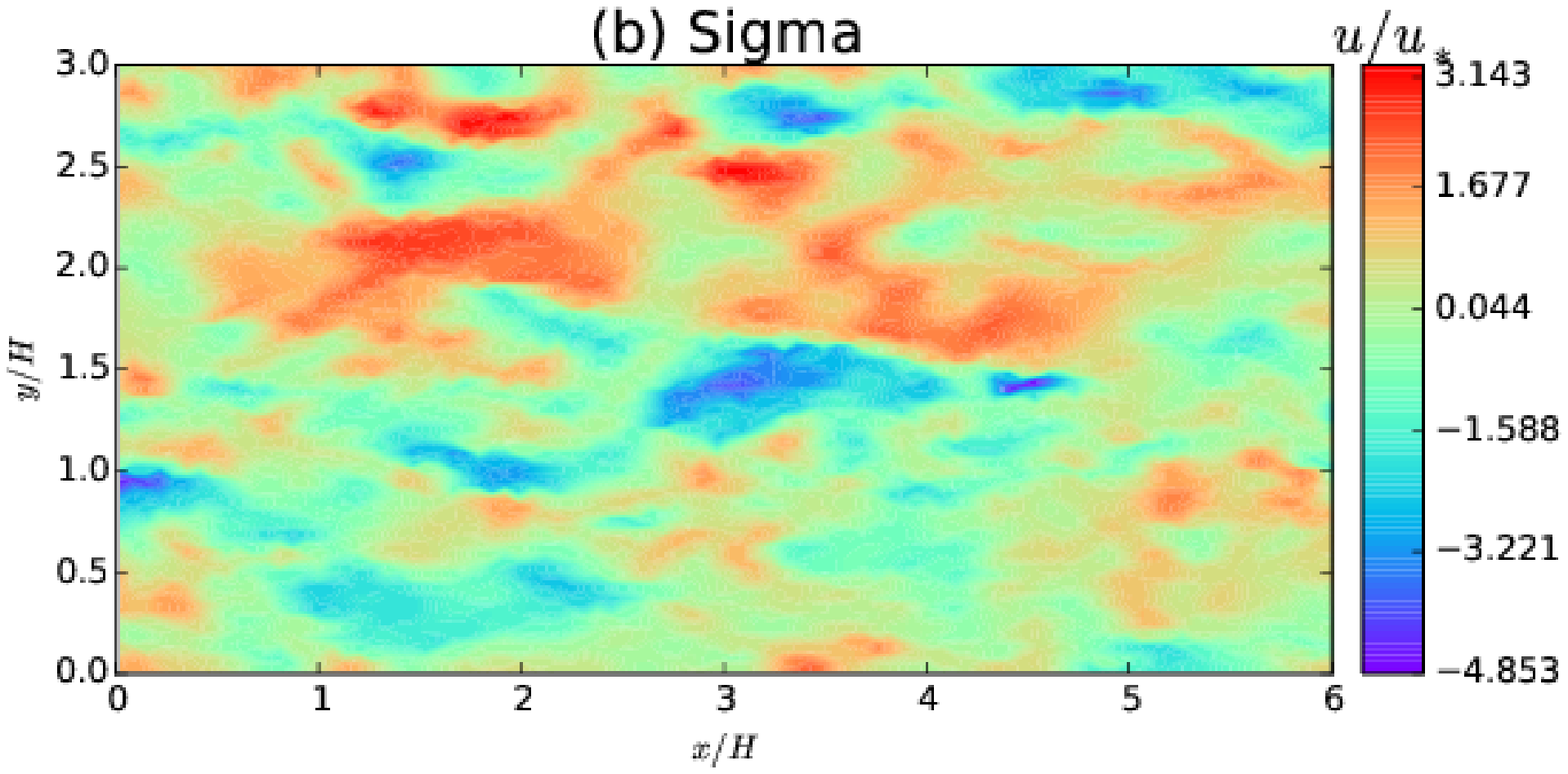}
  \includegraphics[width=0.45\textwidth]{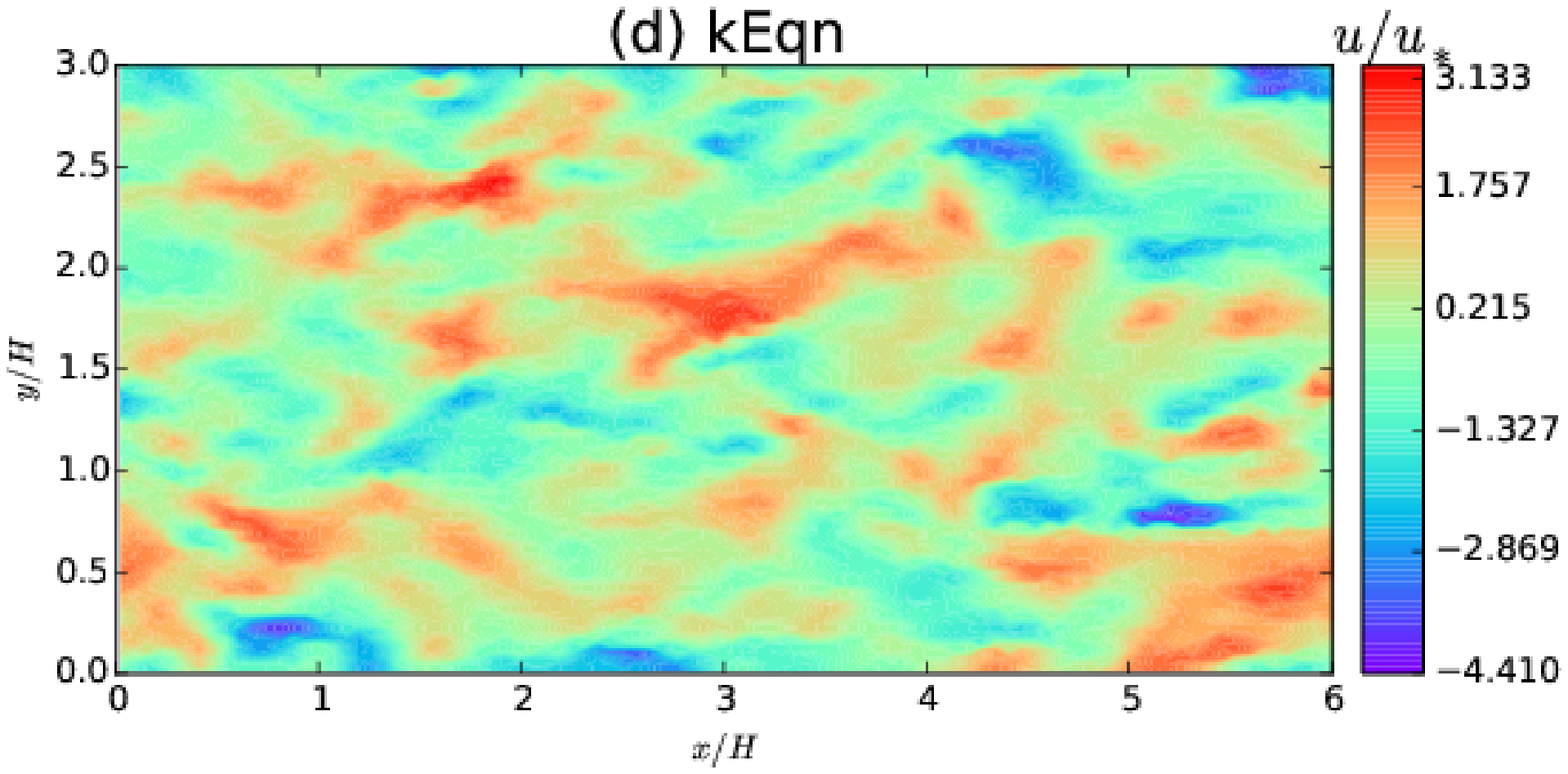}
  \includegraphics[width=0.45\textwidth]{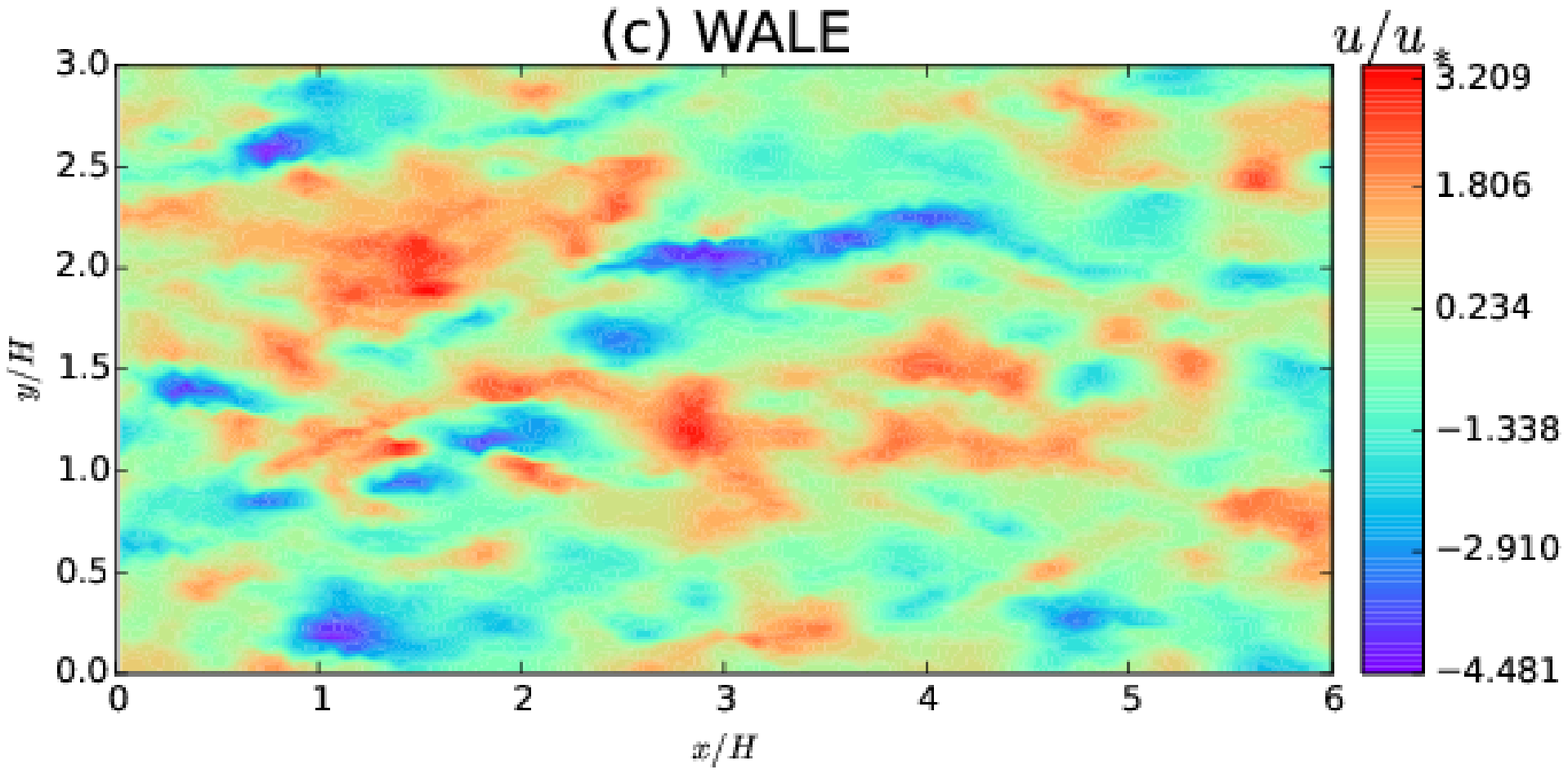}
  \caption{Contour of instantaneous velocity fluctuation at $z/H=0.5$ (in the outer layer). 
    (a) CLES, (b) Sigma, (c) WALE, (d) kEqn. Turbulent structures are similar 
    in the outer layer.}
  \label{fig:contour05}
\end{figure}

Figure~\ref{fig:contour005} and~\ref{fig:contour05} show the fields of instantaneous 
velocity fluctuations at $z/H = 0.05$ (in the buffer layer) and $z/H=0.5$ (in the outer 
layer). It is clearly seen that the resolved turbulent structures are significantly different 
in the buffer layer for different models, while they are undistinguishable in the outer 
layer. In the buffer layer, the turbulence is dominated by the streamwise streaks in all 
models, but the streaks are much finer in Sigma and WALE. The reduced content in 
the near-wall flow structures for kEqn and CSGS may be due to the excessive over-estimation of 
the SGS dissipation, as shown in Fig.~\ref{fig:sgsDissip} (right).

\begin{figure}[!ht]
  \centering
  \includegraphics[width=0.45\textwidth]{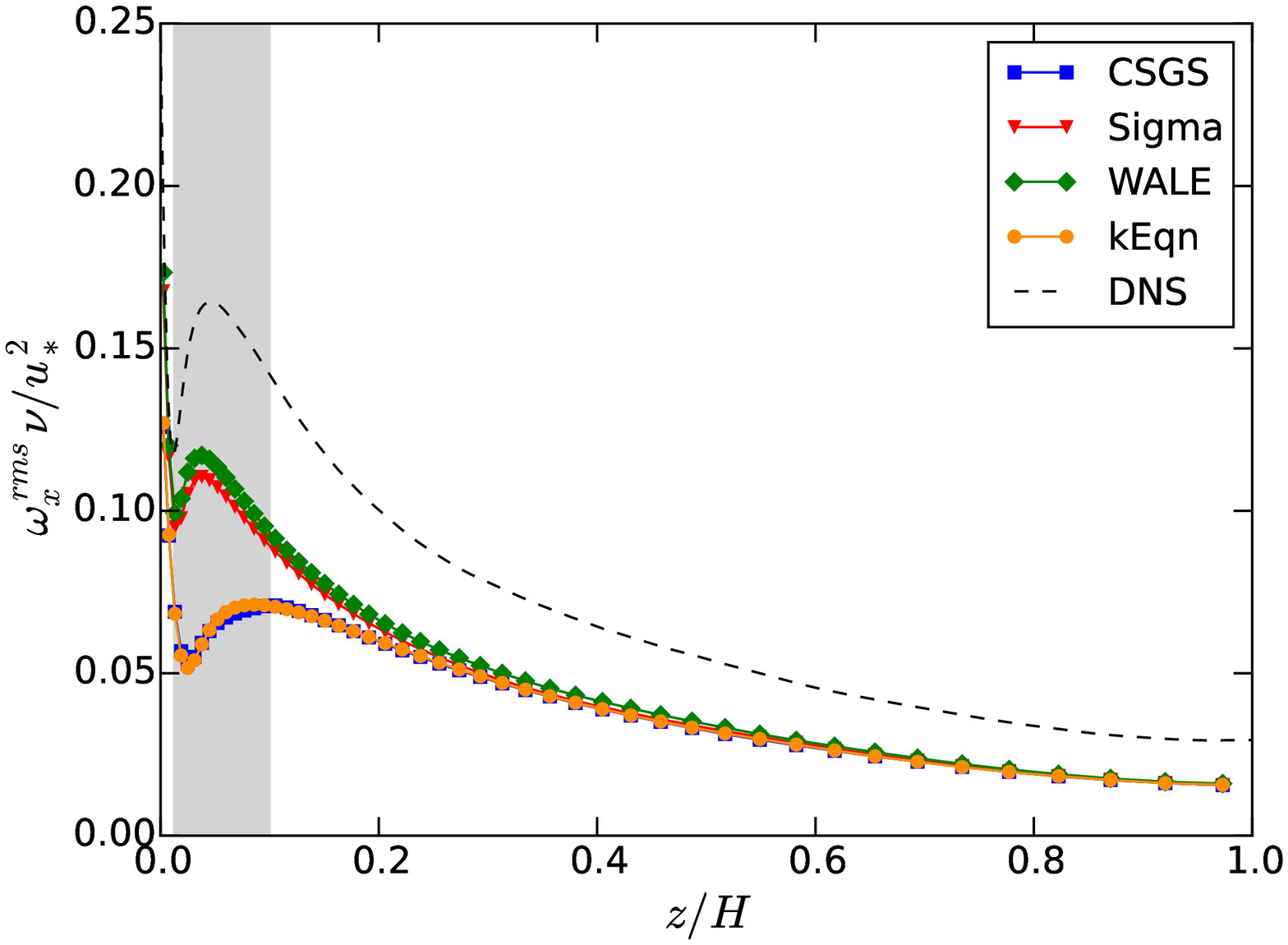}
  \includegraphics[width=0.45\textwidth]{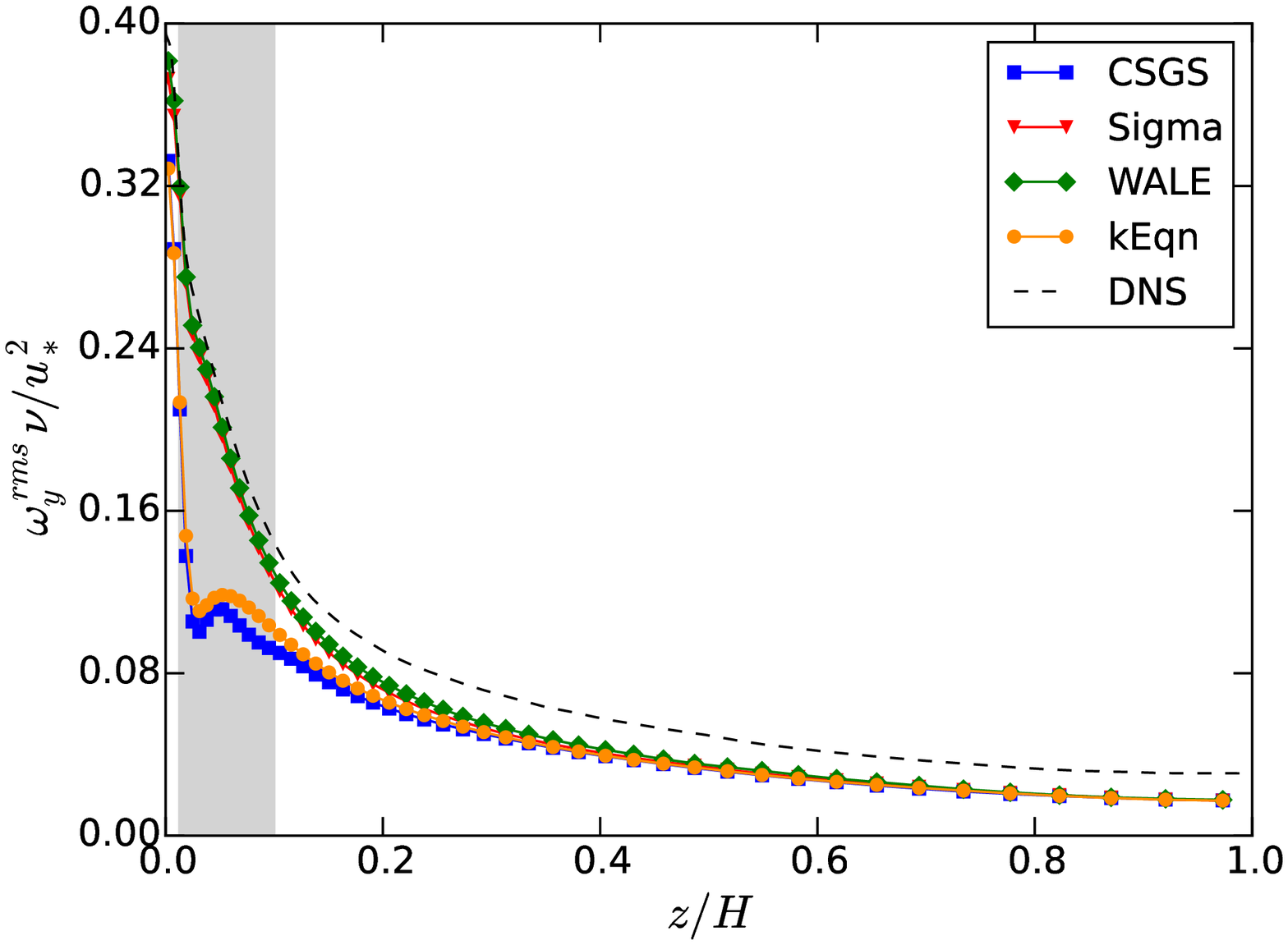}
  \includegraphics[width=0.45\textwidth]{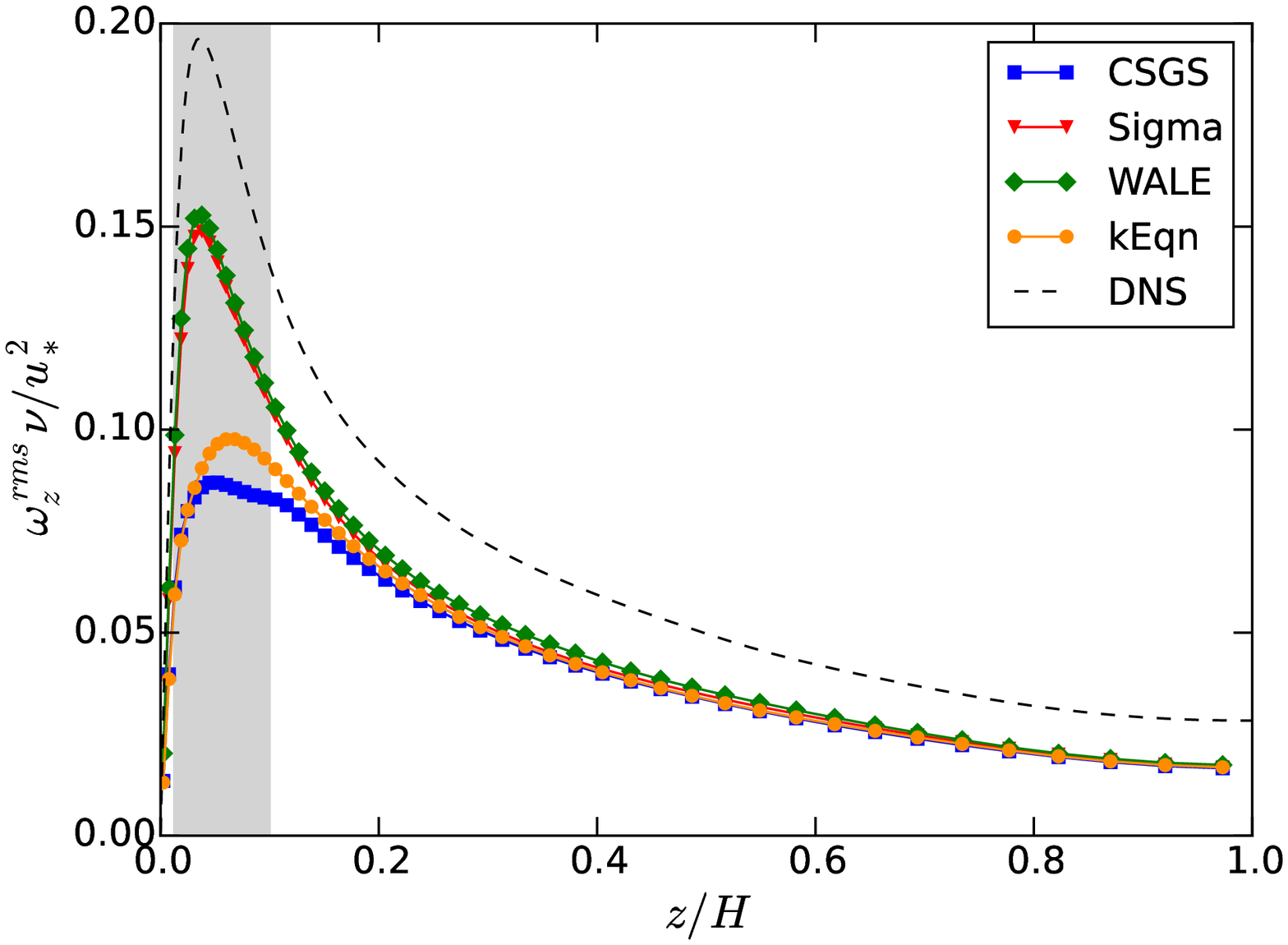}
  \caption{Profiles of the root-mean-square of the fluctuating vorticities.}
  \label{fig:vort}
\end{figure}

The fluctuations of vorticity are next analyzed. 
Vorticity fluctuations are thought to be more 
relevent to small-scale turbulent structures than velocity 
flucturations~\cite{MoinKim_jfm1982}. Therefore, the vorticity fluctuations are 
considered as a better measure for the accuracy of SGS models. 
The profiles of the root-mean-square (RMS) of the vorticity fluctuations are 
shown in Fig.~\ref{fig:vort}. 
The values for the WALE and Sigma model are considerably 
closer to the DNS results than for other models, notably in the buffer layer where the turbulence is 
the most intense. The local minimum 
and maximum in $\omega_x^{rms}$ are found in all models, meaning that the near-wall coherent 
structures (streamwise streaks and vortices) are well captured and are consistent 
with the contour plots in Fig.~\ref{fig:contour005}. The differences are in 
the magnitude of the fluctuations and the location of the local extremum. In the
$\omega_y^{rms}$ profile, the WALE and Sigma model follow very closely 
the DNS data near the wall, while kEqn and CSGS result in an artificial 
local extremum. In $\omega_x^{rms}$ and $\omega_z^{rms}$, WALE and Sigma predict 
up to twice better in the buffer layer than kEqn and CSGS.
In the outer layer, all vorticity components are clearly 
underestimated, whereas for the velocity variances the component $uu$ is 
well resolved by all models, as shown in Fig.~\ref{fig:uu}. 
Again, it is shown that the CSGS model yields even worse results in fluctuations. 

To test the dependence of the results on the location of the constraint interface, 
the interface below which the constraint is imposed, we have performed a group of 
simulations with three interface locations, $z_i = 0.01H, 0.05H, 0.1H$. 
These three locations are in the viscous sublayer, buffer and logarithmic layer, respectively. 
The results are compared with the ones from the simulation without constraint and from the DNS. 
To test whether the baseline model has influence on the constraint model, we employ here 
the WALE model as the baseline model. Other parameters are identical as in the previous 
simulations. 

\begin{figure}[!ht]
  \centering
  \includegraphics[width=0.7\textwidth]{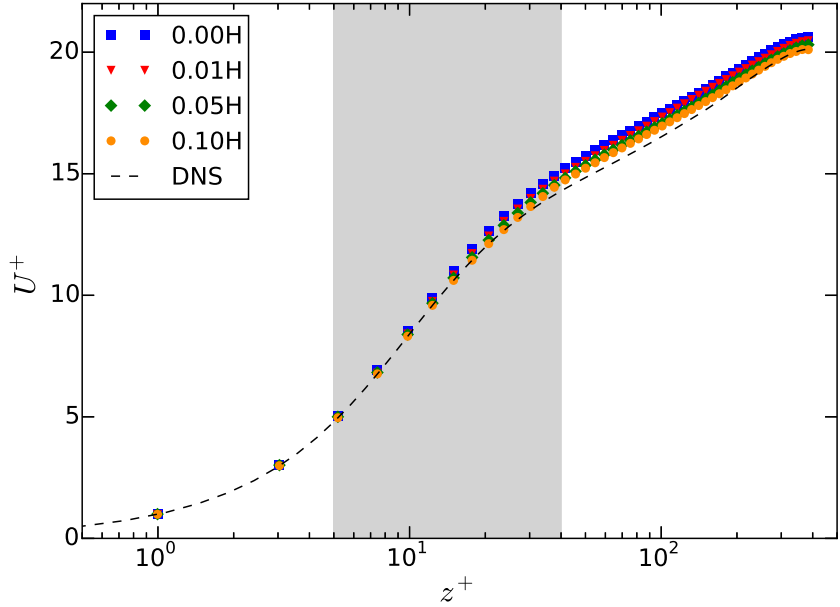}
  \caption{Profiles of the mean streamwise velocity in wall units for different locations 
    of the constraint interface. 
    The dashed line is the DNS data from~\cite{Moser_pof1999}.
    The shaded region is the buffer layer ($z^+ \in [5,40]$ or $z \in [0.013,0.1]H$).
    The legend denotes the location of the constraint interface, below which 
    the constraint is on. 
  }
  \label{fig:U_appx}
\end{figure}

\begin{figure}[!ht]
  \centering
  \includegraphics[width=0.45\textwidth]{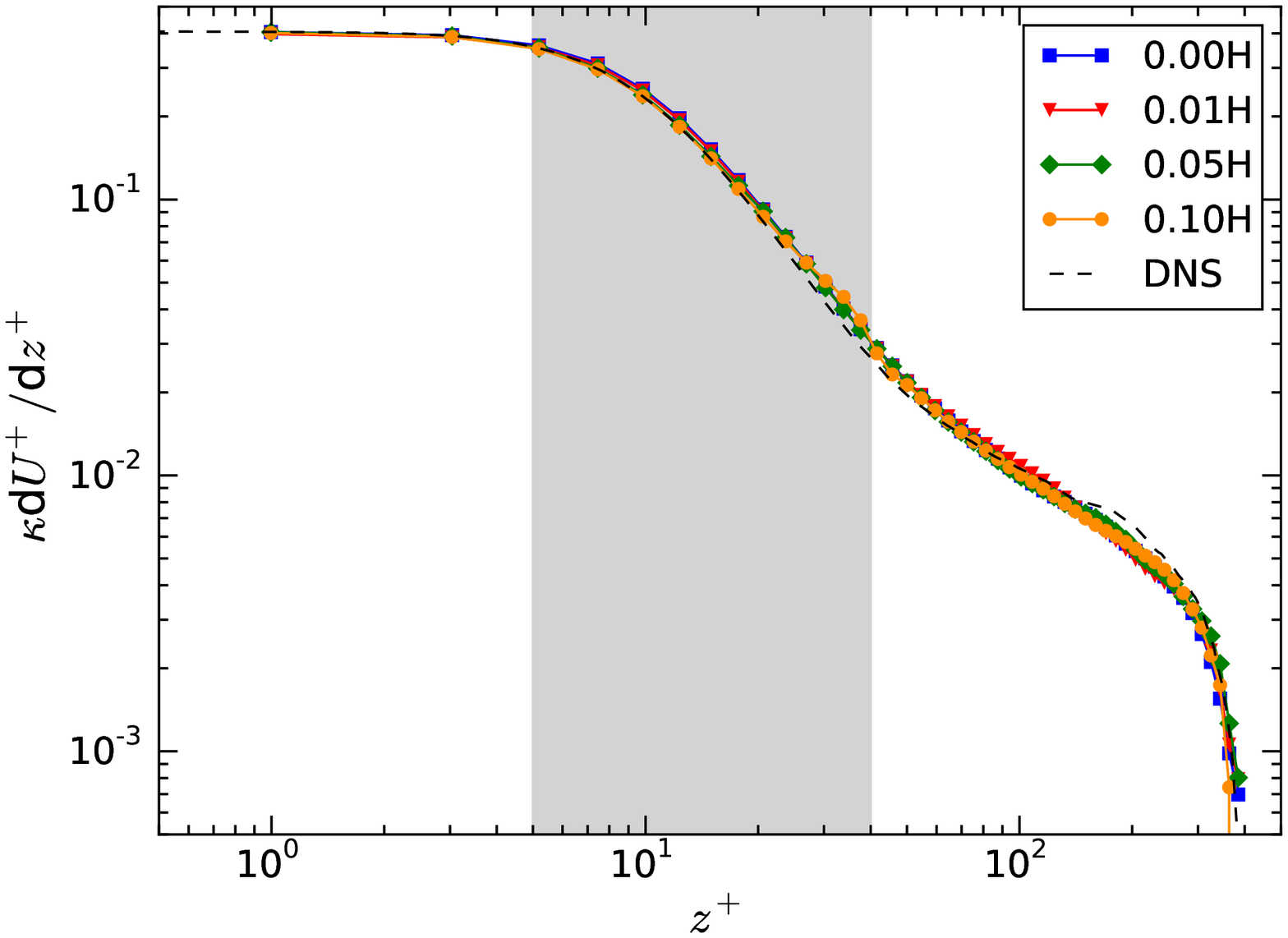}
  \includegraphics[width=0.45\textwidth]{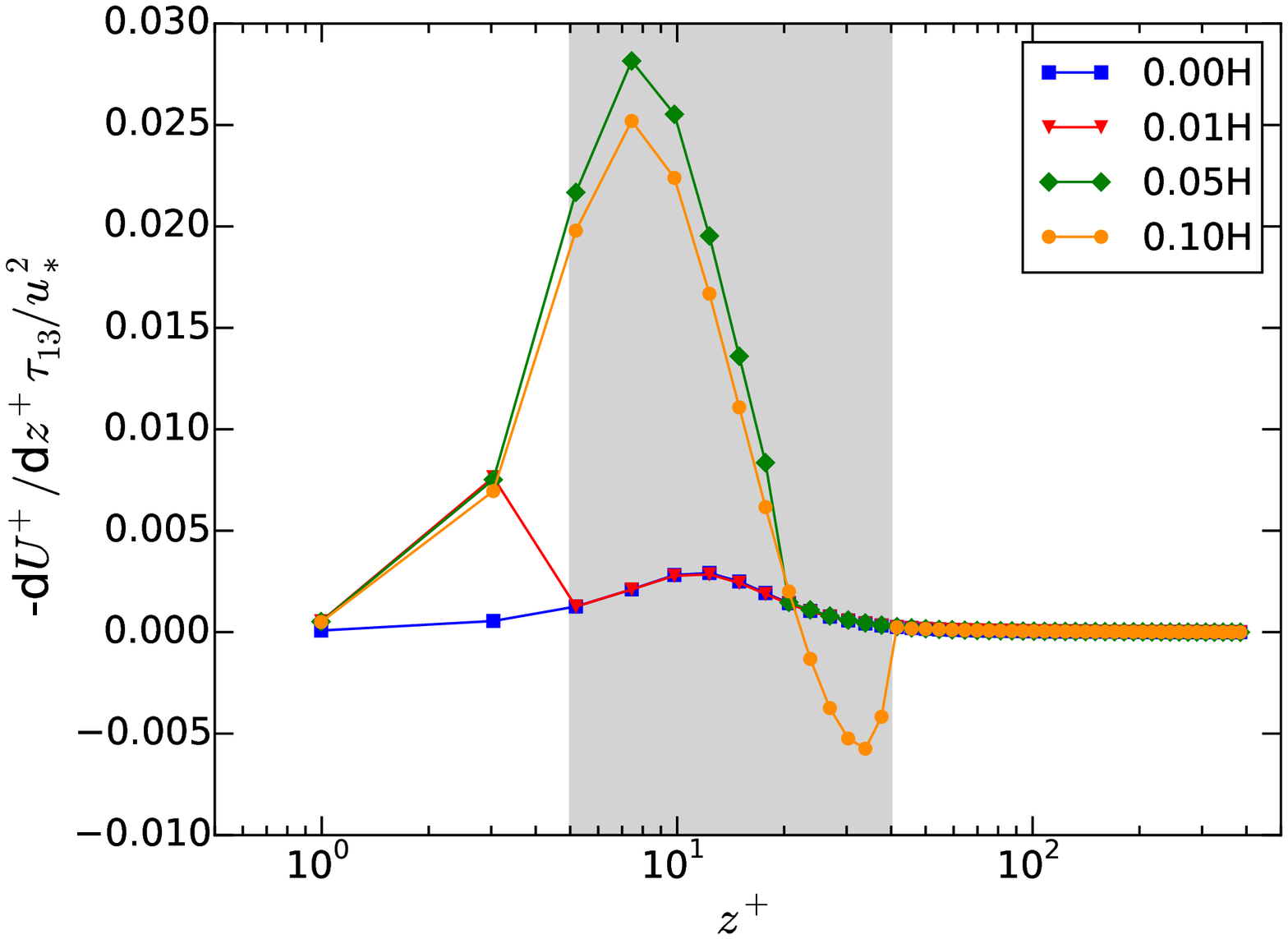}
  \caption{Profiles of the normalized velocity gradient (left) and the normalized mean-flow SGS 
    dissipation $\tau_{13}\text{d}U/\text{d}z$ (right).}
  \label{fig:sgsDissip_appx}
\end{figure}

Figure~\ref{fig:U_appx} shows the mean streamwise velocity profiles\added{ for different constraint interface locations}. 
It is shown that 
as the constraint interface moves away from the wall, the difference between the log layer and 
the DNS results becomes smaller. For $z_i = 0.01H$, the profile is almost the same as the 
one from the simulation without constraint. These observations are in agreement with \added{Chen et al.}~\cite{ChenShi_jfm2012}, 
\changed{even though the simulations in~\cite{ChenShi_jfm2012} are conducted with a spectral code}
{regardless of the pseudo-spectral method used in the simulations in the latter}. 
However, the velocity gradient (Fig.~\ref{fig:sgsDissip_appx} left), the second moments and the 
fluctuating vorticities display negligible difference for different interface locations. 
The interface location, nevertheless, changes significantly the mean-flow SGS dissipation 
in the near-wall region, especially the buffer layer, as shown in 
Fig.~\ref{fig:sgsDissip_appx} (right). Here for WALE, the constraint increases the 
SGS dissipation in the buffer layer, whereas, for kEqn, it decreases the SGS dissipation 
(Fig.~\ref{fig:sgsDissip} right). This provides strong evidence that in the buffer layer, 
the kEqn model overpredict the SGS dissipation while WALE results in an under-estimation.

\section{Discussion and conclusions} \label{sec:conclusion}

In this paper a comparative study of various SGS models 
was conducted by performing a compaign of OpenFOAM finite-volume wall-resolved 
LES simulations. 
The canonical channel flow at $Re_{\tau}=395$ was chosen as the working flow and the DNS results 
from Moser et al.~\cite{Moser_pof1999} as a reference. 
Four SGS models were studied, including kEqn, WALE, Sigma and CSGS. 
The geometry was the same as in the paper by Moser et al.~\cite{Moser_pof1999}, $(2\pi \times \pi \times 2)H$, 
with a resolution of $(96 \times 96 \times 96)$ and $z^+ \simeq 1$ for the first point. 
A constant pressure gradient was employed as the driving force of 
the flow. The CSGS and Sigma model, together with the source term, 
were implemented in OpenFOAM 3.0.1. The new implementation strategy for the CSGS model 
can be easily incorporated into the existing LES codes. 

By analyzing the turbulence statistics up to second order, significant differences 
were found in the SGS models, especially in the buffer layers where 
the turbulence intensities were the strongest. The WALE and Sigma models result in 
much stronger velocity and vorticity fluctuations, as well as much finer flow structures, 
near the wall. 
This may be explained by the significantly smaller SGS viscosity and 
the resultant smaller SGS dissipation inherent in these two models. Compared to its 
baseline model kEqn, the CSGS with constraints on the mean Reynolds shear stress 
had a marginal improvement in the mean velocity but resulted in a slightly worse 
prediction on the fluctuations. Compared to the good results achieved with 
a spectral code in~\cite{ChenShi_jfm2012}, 
the deteriorated performance of CSGS here is probably due 
to the finite volume method used in the simulations, 
which is also consistent with the observations by Verma et al.~\cite{VermaMahesh_pof2013}. 
This seems consistent with the arguments in~\cite{Meneveau_pof1994} that 
ensuring correct Reynolds 
shear stresses is only a necessary but not a sufficient condition for yielding accurate 
statistics. Overall, these differences
reveal the crucial role of the SGS models in the buffer layer in 
wall-resolved LES simulations and more generally in the near-wall regions.

The SGS models considered in the paper also share common features. 
First, in the log layer and 
outer layer, all models resulted in a mean velocity profile which agrees reasonably well with 
the DNS results. A slight log-layer mismatch is present in each model. Different from 
the wall-modelled simulations in previous 
studies~\cite{PorteAgel_jfm2000, KawaiLarsson_pof2012,WuMeyers_pof2013}, 
the mismatch here was shown to have little correlation with the SGS viscosity and dissipation.
The underlying mechanisms for the mismatch are much more complicated. 
Second, most turbulence statistics are underestimated compared to the DNS data.  
This systematic underestimation is likely a result of the lesser resolutions, 
but may also be attributed to a fundamental problem intrinsic in the 
eddy-viscosity family, i.e., the assumption that the SGS stress is linearly proportional 
to the rate-of-strain tensor is incorrect~\cite{MeneveauKatz_annRevFluidMech2000}. 
Nonlinear gradient models had been proprosed~\cite{LiuKatz_jfm1994} but they were reported to 
achieve only a marginal improvement. 
Another important missing component in the current majority 
SGS models is stochasticity. Although governed by deterministic equations, turbulence is 
stochastic. Previous studies
~\cite{MarstorpJohansson_pof2007,Frederiksen_ptrsa2010,RasamJohansson_pof2014} 
already showed in certain cases that stochastic effects improve the fluctuation magnitudes 
and their anisotropy. Hence, building appropriate practical stochastic SGS models is desired to 
adequately capture the random backscattering of the SGS flow motions. 

A constantly ignored quantity in the study of SGS models is the pressure fluctuation,  
which is often considered as less relevant to the flow dynamics. 
As Pope indicated in~\cite{Pope_turbFlow2000}, 
\textit{``the primary effect of the fluctuating pressure is to redistribute 
the energy among the components -- to extract energy from 
$\langle uu\rangle$ and transfer it to $\langle vv\rangle$ and $\langle ww\rangle$.''} 
However, the interaction between the SGS model and the resolved pressure fluctuation 
is unknown. More than 30 years ago, 
Moin and \changed{King}{Kim}~\cite{MoinKim_jfm1982} had already speculated that an appreciable 
portion of the pressure fluctuation may reside in SGS motions and that 
\textit{``the splatting effect is an important property of the flow in the vicinity of the 
walls and should be taken into account in the modeling of near-wall turbulence.''} 
Different from the ``primary'' effect on the bulk flow, the splatting effect denotes the 
transfer of energy from $\langle ww\rangle$ to the other two components 
due to the presence of the wall. 
These two types of effects are seldom considered in most SGS models. 
According to the results from WALE and Sigma models, 
the energy from $\langle uu\rangle$ seems to be incorrectly 
transferred to the other two components,
resulting in an overprediction of $\langle uu\rangle$. 
This transfer of energy is mainly affected through the pressure-rate-of-strain tensor 
$\langle pS_{ij}\rangle$.
To the author's best knowledge, no DNS or experimental data are available for 
this tensor, making it difficult to check the above speculation. 
Nevertheless, the discovery of a relationship between the 
pressure-rate-of-strain tensor and the 
velocity-accelaration correlation~\cite{Pope_jfm2014} offers a feasible way for 
the experimentalists to gain insights into this quantity and to check whether 
the normal SGS stress should be directly modeled.

\section*{Acknowledgement}
The author wish to thank Dr. DongHun Yeo and Dr. Emil Simiu
of the National Institute of Standards and Technology (NIST), 
who provided helpful comments on this work. 
The financial support from the NIST Director Postdoctoral Fellowship is 
acknowledged. Special thanks are due to Mr. Paul Dickey at the Engineering Laboratory 
System Administration (ELSA) for his excellent technical support. 

\pagebreak
\bibliography{./ref_abl.bib}

\begin{thebibliography}{34}%
\makeatletter
\providecommand \@ifxundefined [1]{%
 \@ifx{#1\undefined}
}%
\providecommand \@ifnum [1]{%
 \ifnum #1\expandafter \@firstoftwo
 \else \expandafter \@secondoftwo
 \fi
}%
\providecommand \@ifx [1]{%
 \ifx #1\expandafter \@firstoftwo
 \else \expandafter \@secondoftwo
 \fi
}%
\providecommand \natexlab [1]{#1}%
\providecommand \enquote  [1]{``#1''}%
\providecommand \bibnamefont  [1]{#1}%
\providecommand \bibfnamefont [1]{#1}%
\providecommand \citenamefont [1]{#1}%
\providecommand \href@noop [0]{\@secondoftwo}%
\providecommand \href [0]{\begingroup \@sanitize@url \@href}%
\providecommand \@href[1]{\@@startlink{#1}\@@href}%
\providecommand \@@href[1]{\endgroup#1\@@endlink}%
\providecommand \@sanitize@url [0]{\catcode `\\12\catcode `\$12\catcode
  `\&12\catcode `\#12\catcode `\^12\catcode `\_12\catcode `\%12\relax}%
\providecommand \@@startlink[1]{}%
\providecommand \@@endlink[0]{}%
\providecommand \url  [0]{\begingroup\@sanitize@url \@url }%
\providecommand \@url [1]{\endgroup\@href {#1}{\urlprefix }}%
\providecommand \urlprefix  [0]{URL }%
\providecommand \Eprint [0]{\href }%
\providecommand \doibase [0]{http://dx.doi.org/}%
\providecommand \selectlanguage [0]{\@gobble}%
\providecommand \bibinfo  [0]{\@secondoftwo}%
\providecommand \bibfield  [0]{\@secondoftwo}%
\providecommand \translation [1]{[#1]}%
\providecommand \BibitemOpen [0]{}%
\providecommand \bibitemStop [0]{}%
\providecommand \bibitemNoStop [0]{.\EOS\space}%
\providecommand \EOS [0]{\spacefactor3000\relax}%
\providecommand \BibitemShut  [1]{\csname bibitem#1\endcsname}%
\let\auto@bib@innerbib\@empty
\bibitem [{\citenamefont {Lesieur}\ and\ \citenamefont
  {M\'etais}(1996)}]{LesieurMetais_annRevFluidMech1996}%
  \BibitemOpen
  \bibfield  {author} {\bibinfo {author} {\bibfnamefont {M.}~\bibnamefont
  {Lesieur}}\ and\ \bibinfo {author} {\bibfnamefont {O.}~\bibnamefont
  {M\'etais}},\ }\bibfield  {title} {\enquote {\bibinfo {title} {New trends in
  large-eddy simulations of turbulence},}\ }\href@noop {} {\bibfield  {journal}
  {\bibinfo  {journal} {Annu.~Rev.~Fluid~Mech.}\ }\textbf {\bibinfo {volume}
  {28}},\ \bibinfo {pages} {45--82} (\bibinfo {year} {1996})}\BibitemShut
  {NoStop}%
\bibitem [{\citenamefont {Meneveau}\ and\ \citenamefont
  {Katz}(2000)}]{MeneveauKatz_annRevFluidMech2000}%
  \BibitemOpen
  \bibfield  {author} {\bibinfo {author} {\bibfnamefont {C.}~\bibnamefont
  {Meneveau}}\ and\ \bibinfo {author} {\bibfnamefont {J.}~\bibnamefont
  {Katz}},\ }\bibfield  {title} {\enquote {\bibinfo {title} {Scale-invariance
  and turbulence models for large-eddy simulation},}\ }\href@noop {} {\bibfield
   {journal} {\bibinfo  {journal} {Annu.~Rev.~Fluid~Mech.}\ }\textbf {\bibinfo
  {volume} {32}},\ \bibinfo {pages} {1--32} (\bibinfo {year}
  {2000})}\BibitemShut {NoStop}%
\bibitem [{\citenamefont {Smagorinsky}(1963)}]{Smagorinsky_MonWeatherRev1963}%
  \BibitemOpen
  \bibfield  {author} {\bibinfo {author} {\bibfnamefont {J.}~\bibnamefont
  {Smagorinsky}},\ }\bibfield  {title} {\enquote {\bibinfo {title} {General
  circulation experiments with the primitive equations. {I}. the basic
  experiment},}\ }\href@noop {} {\bibfield  {journal} {\bibinfo  {journal}
  {Mon. Weather Rev.}\ }\textbf {\bibinfo {volume} {91}},\ \bibinfo {pages}
  {99--164} (\bibinfo {year} {1963})}\BibitemShut {NoStop}%
\bibitem [{\citenamefont {Lilly}(1967)}]{Lilly_ibm1967}%
  \BibitemOpen
  \bibfield  {author} {\bibinfo {author} {\bibfnamefont {D.~K.}\ \bibnamefont
  {Lilly}},\ }\bibfield  {title} {\enquote {\bibinfo {title} {The
  representation of small-scale turbulence in numerical simulation
  experiments},}\ }\href@noop {} {\bibfield  {journal} {\bibinfo  {journal}
  {Proc. IBM Scientific Computing Symp. Environ. Sci.}\ ,\ \bibinfo {pages}
  {195}} (\bibinfo {year} {1967})}\BibitemShut {NoStop}%
\bibitem [{\citenamefont {Schumann}(1975)}]{Schumann_jcp1975}%
  \BibitemOpen
  \bibfield  {author} {\bibinfo {author} {\bibfnamefont {U.}~\bibnamefont
  {Schumann}},\ }\bibfield  {title} {\enquote {\bibinfo {title} {Subgrid scale
  model for finite difference simulations of turbulent flows in plane channels
  and annuli},}\ }\href@noop {} {\bibfield  {journal} {\bibinfo  {journal}
  {J.~Comput.~Phys.}\ }\textbf {\bibinfo {volume} {18}},\ \bibinfo {pages}
  {376--404} (\bibinfo {year} {1975})}\BibitemShut {NoStop}%
\bibitem [{\citenamefont {Germano}\ \emph {et~al.}(1991)\citenamefont
  {Germano}, \citenamefont {Piomelli}, \citenamefont {Moin},\ and\
  \citenamefont {Cabot}}]{GermanoCabot_pof1991}%
  \BibitemOpen
  \bibfield  {author} {\bibinfo {author} {\bibfnamefont {M.}~\bibnamefont
  {Germano}}, \bibinfo {author} {\bibfnamefont {U.}~\bibnamefont {Piomelli}},
  \bibinfo {author} {\bibfnamefont {P.}~\bibnamefont {Moin}}, \ and\ \bibinfo
  {author} {\bibfnamefont {W.~H.}\ \bibnamefont {Cabot}},\ }\bibfield  {title}
  {\enquote {\bibinfo {title} {A dynamic subgrid-scale eddy viscosity model},}\
  }\href@noop {} {\bibfield  {journal} {\bibinfo  {journal} {Phys.~Fluids}\
  }\textbf {\bibinfo {volume} {A3}},\ \bibinfo {pages} {1760--1765} (\bibinfo
  {year} {1991})}\BibitemShut {NoStop}%
\bibitem [{\citenamefont {Port\'e-Agel}, \citenamefont {Meneveau},\ and\
  \citenamefont {Parlange}(2000)}]{PorteAgel_jfm2000}%
  \BibitemOpen
  \bibfield  {author} {\bibinfo {author} {\bibfnamefont {F.}~\bibnamefont
  {Port\'e-Agel}}, \bibinfo {author} {\bibfnamefont {C.}~\bibnamefont
  {Meneveau}}, \ and\ \bibinfo {author} {\bibfnamefont {M.~B.}\ \bibnamefont
  {Parlange}},\ }\bibfield  {title} {\enquote {\bibinfo {title} {A
  scale-dependent dynamic model for large-eddy simulation: application to a
  neutral atmospheric boundary layer},}\ }\href@noop {} {\bibfield  {journal}
  {\bibinfo  {journal} {J.~Fluid~Mech.}\ }\textbf {\bibinfo {volume} {415}},\
  \bibinfo {pages} {261--284} (\bibinfo {year} {2000})}\BibitemShut {NoStop}%
\bibitem [{\citenamefont {Trias}\ \emph {et~al.}(2015)\citenamefont {Trias},
  \citenamefont {Folch}, \citenamefont {Gorobets},\ and\ \citenamefont
  {Oliva}}]{TriasOliva_pof2015}%
  \BibitemOpen
  \bibfield  {author} {\bibinfo {author} {\bibfnamefont {F.~X.}\ \bibnamefont
  {Trias}}, \bibinfo {author} {\bibfnamefont {D.}~\bibnamefont {Folch}},
  \bibinfo {author} {\bibfnamefont {A.}~\bibnamefont {Gorobets}}, \ and\
  \bibinfo {author} {\bibfnamefont {A.}~\bibnamefont {Oliva}},\ }\bibfield
  {title} {\enquote {\bibinfo {title} {Building proper invariants for
  eddy-viscosity subgrid-scale models},}\ }\href@noop {} {\bibfield  {journal}
  {\bibinfo  {journal} {Phys.~Fluids}\ }\textbf {\bibinfo {volume} {27}},\
  \bibinfo {pages} {065103} (\bibinfo {year} {2015})}\BibitemShut {NoStop}%
\bibitem [{\citenamefont {Nicoud}\ and\ \citenamefont
  {Ducros}(1999)}]{NicoudDucros_1999}%
  \BibitemOpen
  \bibfield  {author} {\bibinfo {author} {\bibfnamefont {F.}~\bibnamefont
  {Nicoud}}\ and\ \bibinfo {author} {\bibfnamefont {F.}~\bibnamefont
  {Ducros}},\ }\bibfield  {title} {\enquote {\bibinfo {title} {Subgrid-scale
  stress modelling based on the square of the velocity gradient tensor},}\
  }\href@noop {} {\bibfield  {journal} {\bibinfo  {journal} {FLow, Turbulence
  and Combustion}\ }\textbf {\bibinfo {volume} {62}},\ \bibinfo {pages}
  {183--200} (\bibinfo {year} {1999})}\BibitemShut {NoStop}%
\bibitem [{\citenamefont {Vreman}(2004)}]{Vreman_pof2004}%
  \BibitemOpen
  \bibfield  {author} {\bibinfo {author} {\bibfnamefont {A.~W.}\ \bibnamefont
  {Vreman}},\ }\bibfield  {title} {\enquote {\bibinfo {title} {An
  eddy-viscosity subgrid-scale model for turbulent shear flow: Algebraic theory
  and applications},}\ }\href@noop {} {\bibfield  {journal} {\bibinfo
  {journal} {Phys.~Fluids}\ }\textbf {\bibinfo {volume} {16}},\ \bibinfo
  {pages} {3670--3681} (\bibinfo {year} {2004})}\BibitemShut {NoStop}%
\bibitem [{\citenamefont {Verstappen}(2011)}]{Verstappen_jsc2011}%
  \BibitemOpen
  \bibfield  {author} {\bibinfo {author} {\bibfnamefont {R.}~\bibnamefont
  {Verstappen}},\ }\bibfield  {title} {\enquote {\bibinfo {title} {When does
  eddy viscosity damp subfilter scales sufficiently?}}\ }\href@noop {}
  {\bibfield  {journal} {\bibinfo  {journal} {J. Sci. Comput.}\ }\textbf
  {\bibinfo {volume} {49}},\ \bibinfo {pages} {94--110} (\bibinfo {year}
  {2011})}\BibitemShut {NoStop}%
\bibitem [{\citenamefont {Nicoud}\ \emph {et~al.}(2011)\citenamefont {Nicoud},
  \citenamefont {Toda}, \citenamefont {Cabrit}, \citenamefont {Bose},\ and\
  \citenamefont {Lee}}]{NicoudLee_pof2011}%
  \BibitemOpen
  \bibfield  {author} {\bibinfo {author} {\bibfnamefont {F.}~\bibnamefont
  {Nicoud}}, \bibinfo {author} {\bibfnamefont {H.~B.}\ \bibnamefont {Toda}},
  \bibinfo {author} {\bibfnamefont {O.}~\bibnamefont {Cabrit}}, \bibinfo
  {author} {\bibfnamefont {S.}~\bibnamefont {Bose}}, \ and\ \bibinfo {author}
  {\bibfnamefont {J.}~\bibnamefont {Lee}},\ }\bibfield  {title} {\enquote
  {\bibinfo {title} {Using singular values to build a subgrid-scale model for
  large eddy simulations},}\ }\href@noop {} {\bibfield  {journal} {\bibinfo
  {journal} {Phys.~Fluids}\ }\textbf {\bibinfo {volume} {23}},\ \bibinfo
  {pages} {085106} (\bibinfo {year} {2011})}\BibitemShut {NoStop}%
\bibitem [{\citenamefont {Sullivan}, \citenamefont {McWilliams},\ and\
  \citenamefont {Moeng}(1994)}]{SullivanMoeng_blm1994}%
  \BibitemOpen
  \bibfield  {author} {\bibinfo {author} {\bibfnamefont {P.~P.}\ \bibnamefont
  {Sullivan}}, \bibinfo {author} {\bibfnamefont {J.~C.}\ \bibnamefont
  {McWilliams}}, \ and\ \bibinfo {author} {\bibfnamefont {C.-H.}\ \bibnamefont
  {Moeng}},\ }\bibfield  {title} {\enquote {\bibinfo {title} {A subgrid-scale
  model for large-eddy simulation of planetary boundary-layer flows},}\
  }\href@noop {} {\bibfield  {journal} {\bibinfo  {journal}
  {Boundary-Layer~Meteorol.}\ }\textbf {\bibinfo {volume} {71}},\ \bibinfo
  {pages} {247--276} (\bibinfo {year} {1994})}\BibitemShut {NoStop}%
\bibitem [{\citenamefont {L\'ev\^eque}\ \emph {et~al.}(2007)\citenamefont
  {L\'ev\^eque}, \citenamefont {Toshi}, \citenamefont {Shao},\ and\
  \citenamefont {Bertoglio}}]{LevequeBertoglio_jfm2007}%
  \BibitemOpen
  \bibfield  {author} {\bibinfo {author} {\bibfnamefont {E.}~\bibnamefont
  {L\'ev\^eque}}, \bibinfo {author} {\bibfnamefont {F.}~\bibnamefont {Toshi}},
  \bibinfo {author} {\bibfnamefont {L.}~\bibnamefont {Shao}}, \ and\ \bibinfo
  {author} {\bibfnamefont {J.-P.}\ \bibnamefont {Bertoglio}},\ }\bibfield
  {title} {\enquote {\bibinfo {title} {Shear-improved smagorinsky model for
  large-eddy simulation of wall-bounded turbulent flows},}\ }\href@noop {}
  {\bibfield  {journal} {\bibinfo  {journal} {J.~Fluid~Mech.}\ }\textbf
  {\bibinfo {volume} {570}},\ \bibinfo {pages} {491--502} (\bibinfo {year}
  {2007})}\BibitemShut {NoStop}%
\bibitem [{\citenamefont {Chen}\ \emph {et~al.}(2012)\citenamefont {Chen},
  \citenamefont {Xia}, \citenamefont {Pei}, \citenamefont {Wang}, \citenamefont
  {Yang}, \citenamefont {Xiao},\ and\ \citenamefont {Shi}}]{ChenShi_jfm2012}%
  \BibitemOpen
  \bibfield  {author} {\bibinfo {author} {\bibfnamefont {S.}~\bibnamefont
  {Chen}}, \bibinfo {author} {\bibfnamefont {Z.}~\bibnamefont {Xia}}, \bibinfo
  {author} {\bibfnamefont {S.}~\bibnamefont {Pei}}, \bibinfo {author}
  {\bibfnamefont {J.}~\bibnamefont {Wang}}, \bibinfo {author} {\bibfnamefont
  {Y.}~\bibnamefont {Yang}}, \bibinfo {author} {\bibfnamefont {Z.}~\bibnamefont
  {Xiao}}, \ and\ \bibinfo {author} {\bibfnamefont {Y.}~\bibnamefont {Shi}},\
  }\bibfield  {title} {\enquote {\bibinfo {title} {Reynolds-stress-constrained
  large-eddy simulation of wall-bounded turbulent flows},}\ }\href@noop {}
  {\bibfield  {journal} {\bibinfo  {journal} {J.~Fluid~Mech.}\ }\textbf
  {\bibinfo {volume} {703}},\ \bibinfo {pages} {1--28} (\bibinfo {year}
  {2012})}\BibitemShut {NoStop}%
\bibitem [{\citenamefont {Yoshizawa}(1985)}]{Yoshizawa_kEq1985}%
  \BibitemOpen
  \bibfield  {author} {\bibinfo {author} {\bibfnamefont {A.}~\bibnamefont
  {Yoshizawa}},\ }\bibfield  {title} {\enquote {\bibinfo {title} {A
  statistically-derived subgrid-scale kinetic energy model for the large-eddy
  simulations of turbulent flows},}\ }\href@noop {} {\bibfield  {journal}
  {\bibinfo  {journal} {Journal of the Physical Society of Japan}\ }\textbf
  {\bibinfo {volume} {54}},\ \bibinfo {pages} {2834--2839} (\bibinfo {year}
  {1985})}\BibitemShut {NoStop}%
\bibitem [{\citenamefont {de~Villiers}(2006)}]{deVille_phdthesis2006}%
  \BibitemOpen
  \bibfield  {author} {\bibinfo {author} {\bibfnamefont {E.}~\bibnamefont
  {de~Villiers}},\ }\emph {\bibinfo {title} {The Potential of Large Eddy
  Simulation for the Modeling of Wall Bounded Flows}},\ \href@noop {} {Ph.D.
  thesis},\ \bibinfo  {school} {Imperial College of Science, Technology and
  Medicine} (\bibinfo {year} {2006})\BibitemShut {NoStop}%
\bibitem [{\citenamefont {Shi}, \citenamefont {Xiao},\ and\ \citenamefont
  {Chen}(2008)}]{ShiChen_pof2008}%
  \BibitemOpen
  \bibfield  {author} {\bibinfo {author} {\bibfnamefont {Y.}~\bibnamefont
  {Shi}}, \bibinfo {author} {\bibfnamefont {Z.}~\bibnamefont {Xiao}}, \ and\
  \bibinfo {author} {\bibfnamefont {S.}~\bibnamefont {Chen}},\ }\bibfield
  {title} {\enquote {\bibinfo {title} {Constrained subgrid-scale stress model
  for large eddy simulation},}\ }\href@noop {} {\bibfield  {journal} {\bibinfo
  {journal} {Phys.~Fluids}\ }\textbf {\bibinfo {volume} {20}},\ \bibinfo
  {pages} {011701} (\bibinfo {year} {2008})}\BibitemShut {NoStop}%
\bibitem [{\citenamefont {Jiang}\ \emph {et~al.}(2013)\citenamefont {Jiang},
  \citenamefont {Xiao}, \citenamefont {Shi},\ and\ \citenamefont
  {Chen}}]{JiangChen_pof2014}%
  \BibitemOpen
  \bibfield  {author} {\bibinfo {author} {\bibfnamefont {Z.}~\bibnamefont
  {Jiang}}, \bibinfo {author} {\bibfnamefont {Z.}~\bibnamefont {Xiao}},
  \bibinfo {author} {\bibfnamefont {Y.}~\bibnamefont {Shi}}, \ and\ \bibinfo
  {author} {\bibfnamefont {S.}~\bibnamefont {Chen}},\ }\bibfield  {title}
  {\enquote {\bibinfo {title} {Constrained large-eddy simulation of
  wall-bounded compressible turbulent flows},}\ }\href@noop {} {\bibfield
  {journal} {\bibinfo  {journal} {Phys.~Fluids}\ }\textbf {\bibinfo {volume}
  {25}},\ \bibinfo {pages} {106102} (\bibinfo {year} {2013})}\BibitemShut
  {NoStop}%
\bibitem [{\citenamefont {Zhao}\ \emph {et~al.}(2014)\citenamefont {Zhao},
  \citenamefont {Xia}, \citenamefont {Shi}, \citenamefont {Xiao},\ and\
  \citenamefont {Chen}}]{ZhaoChen_pof2014}%
  \BibitemOpen
  \bibfield  {author} {\bibinfo {author} {\bibfnamefont {Y.}~\bibnamefont
  {Zhao}}, \bibinfo {author} {\bibfnamefont {Z.}~\bibnamefont {Xia}}, \bibinfo
  {author} {\bibfnamefont {Y.}~\bibnamefont {Shi}}, \bibinfo {author}
  {\bibfnamefont {Z.}~\bibnamefont {Xiao}}, \ and\ \bibinfo {author}
  {\bibfnamefont {S.}~\bibnamefont {Chen}},\ }\bibfield  {title} {\enquote
  {\bibinfo {title} {Constrained large-eddy simulation of laminar-turbulent
  transition in channel flow},}\ }\href@noop {} {\bibfield  {journal} {\bibinfo
   {journal} {Phys.~Fluids}\ }\textbf {\bibinfo {volume} {26}},\ \bibinfo
  {pages} {095103} (\bibinfo {year} {2014})}\BibitemShut {NoStop}%
\bibitem [{\citenamefont {Moser}, \citenamefont {Kim},\ and\ \citenamefont
  {Mansour}(1999)}]{Moser_pof1999}%
  \BibitemOpen
  \bibfield  {author} {\bibinfo {author} {\bibfnamefont {R.~D.}\ \bibnamefont
  {Moser}}, \bibinfo {author} {\bibfnamefont {J.}~\bibnamefont {Kim}}, \ and\
  \bibinfo {author} {\bibfnamefont {N.~N.}\ \bibnamefont {Mansour}},\
  }\bibfield  {title} {\enquote {\bibinfo {title} {Direct numerical simulation
  of turbulent channel flow up to $re_{\tau} = 590$},}\ }\href@noop {}
  {\bibfield  {journal} {\bibinfo  {journal} {Phys.~Fluids}\ }\textbf {\bibinfo
  {volume} {11}},\ \bibinfo {pages} {943} (\bibinfo {year} {1999})}\BibitemShut
  {NoStop}%
\bibitem [{\citenamefont {3.0.1}()}]{openfoam_2015}%
  \BibitemOpen
  \bibfield  {author} {\bibinfo {author} {\bibfnamefont {O.}~\bibnamefont
  {3.0.1}},\ }\href@noop {} {\enquote {\bibinfo {title}
  {http://openfoam.org/release/3-0-1/},}\ }\BibitemShut {NoStop}%
\bibitem [{\citenamefont {Robertson}\ \emph {et~al.}(2015)\citenamefont
  {Robertson}, \citenamefont {Choudhury}, \citenamefont {Bhushan},\ and\
  \citenamefont {Walters}}]{RobertsonWalters_caf2015}%
  \BibitemOpen
  \bibfield  {author} {\bibinfo {author} {\bibfnamefont {E.}~\bibnamefont
  {Robertson}}, \bibinfo {author} {\bibfnamefont {V.}~\bibnamefont
  {Choudhury}}, \bibinfo {author} {\bibfnamefont {S.}~\bibnamefont {Bhushan}},
  \ and\ \bibinfo {author} {\bibfnamefont {D.~K.}\ \bibnamefont {Walters}},\
  }\bibfield  {title} {\enquote {\bibinfo {title} {Validation of {OpenFOAM}
  numerical methods and turbulence models for incompressible bluff body
  flows},}\ }\href@noop {} {\bibfield  {journal} {\bibinfo  {journal}
  {Comput.~Fluids}\ }\textbf {\bibinfo {volume} {123}},\ \bibinfo {pages}
  {122--145} (\bibinfo {year} {2015})}\BibitemShut {NoStop}%
\bibitem [{\citenamefont {Pope}(2000)}]{Pope_turbFlow2000}%
  \BibitemOpen
  \bibfield  {author} {\bibinfo {author} {\bibfnamefont {S.~B.}\ \bibnamefont
  {Pope}},\ }\href@noop {} {\emph {\bibinfo {title} {Turbulent Flows}}}\
  (\bibinfo  {publisher} {Cambridge University Press},\ \bibinfo {year}
  {2000})\BibitemShut {NoStop}%
\bibitem [{\citenamefont {Kawai}\ and\ \citenamefont
  {Larsson}(2012)}]{KawaiLarsson_pof2012}%
  \BibitemOpen
  \bibfield  {author} {\bibinfo {author} {\bibfnamefont {S.}~\bibnamefont
  {Kawai}}\ and\ \bibinfo {author} {\bibfnamefont {J.}~\bibnamefont
  {Larsson}},\ }\bibfield  {title} {\enquote {\bibinfo {title} {Wall-modeling
  in large eddy simulation: Length scales, grid resolution, and accuracy},}\
  }\href@noop {} {\bibfield  {journal} {\bibinfo  {journal} {Phys.~Fluids}\
  }\textbf {\bibinfo {volume} {24}},\ \bibinfo {pages} {015105} (\bibinfo
  {year} {2012})}\BibitemShut {NoStop}%
\bibitem [{\citenamefont {Wu}\ and\ \citenamefont
  {Meyers}(2013)}]{WuMeyers_pof2013}%
  \BibitemOpen
  \bibfield  {author} {\bibinfo {author} {\bibfnamefont {P.}~\bibnamefont
  {Wu}}\ and\ \bibinfo {author} {\bibfnamefont {J.}~\bibnamefont {Meyers}},\
  }\bibfield  {title} {\enquote {\bibinfo {title} {A constraint for the
  subgrid-scale stresses in the logarithmic region of high reynolds number
  turbulent boundary layers: A solution to the log-layer mismatch problem},}\
  }\href@noop {} {\bibfield  {journal} {\bibinfo  {journal} {Phys.~Fluids}\
  }\textbf {\bibinfo {volume} {25}},\ \bibinfo {pages} {015104} (\bibinfo
  {year} {2013})}\BibitemShut {NoStop}%
\bibitem [{\citenamefont {Moin}\ and\ \citenamefont
  {Kim}(1982)}]{MoinKim_jfm1982}%
  \BibitemOpen
  \bibfield  {author} {\bibinfo {author} {\bibfnamefont {P.}~\bibnamefont
  {Moin}}\ and\ \bibinfo {author} {\bibfnamefont {J.}~\bibnamefont {Kim}},\
  }\bibfield  {title} {\enquote {\bibinfo {title} {Numerical investigation of
  turbulent channel flow},}\ }\href@noop {} {\bibfield  {journal} {\bibinfo
  {journal} {J.~Fluid~Mech.}\ }\textbf {\bibinfo {volume} {118}},\ \bibinfo
  {pages} {341--377} (\bibinfo {year} {1982})}\BibitemShut {NoStop}%
\bibitem [{\citenamefont {Verma}, \citenamefont {Park},\ and\ \citenamefont
  {Mahesh}(2013)}]{VermaMahesh_pof2013}%
  \BibitemOpen
  \bibfield  {author} {\bibinfo {author} {\bibfnamefont {A.}~\bibnamefont
  {Verma}}, \bibinfo {author} {\bibfnamefont {N.}~\bibnamefont {Park}}, \ and\
  \bibinfo {author} {\bibfnamefont {K.}~\bibnamefont {Mahesh}},\ }\bibfield
  {title} {\enquote {\bibinfo {title} {A hybrid subgrid-scale model constrained
  by reynolds stress},}\ }\href@noop {} {\bibfield  {journal} {\bibinfo
  {journal} {Phys.~Fluids}\ }\textbf {\bibinfo {volume} {25}},\ \bibinfo
  {pages} {110805} (\bibinfo {year} {2013})}\BibitemShut {NoStop}%
\bibitem [{\citenamefont {Meneveau}(1994)}]{Meneveau_pof1994}%
  \BibitemOpen
  \bibfield  {author} {\bibinfo {author} {\bibfnamefont {C.}~\bibnamefont
  {Meneveau}},\ }\bibfield  {title} {\enquote {\bibinfo {title} {Statistics of
  turbulence subgrid-scale stresses: Necessary conditions and experimental
  tests},}\ }\href@noop {} {\bibfield  {journal} {\bibinfo  {journal}
  {Phys.~Fluids}\ }\textbf {\bibinfo {volume} {6}},\ \bibinfo {pages}
  {815--833} (\bibinfo {year} {1994})}\BibitemShut {NoStop}%
\bibitem [{\citenamefont {Liu}, \citenamefont {Meneveau},\ and\ \citenamefont
  {Katz}(1994)}]{LiuKatz_jfm1994}%
  \BibitemOpen
  \bibfield  {author} {\bibinfo {author} {\bibfnamefont {S.}~\bibnamefont
  {Liu}}, \bibinfo {author} {\bibfnamefont {C.}~\bibnamefont {Meneveau}}, \
  and\ \bibinfo {author} {\bibfnamefont {J.}~\bibnamefont {Katz}},\ }\bibfield
  {title} {\enquote {\bibinfo {title} {On the properties of similarity
  subgrid-scale models as deduced from measurements in a turbulent jet},}\
  }\href@noop {} {\bibfield  {journal} {\bibinfo  {journal} {J.~Fluid~Mech.}\
  }\textbf {\bibinfo {volume} {275}},\ \bibinfo {pages} {83--119} (\bibinfo
  {year} {1994})}\BibitemShut {NoStop}%
\bibitem [{\citenamefont {Marstorp}, \citenamefont {Brethouwer},\ and\
  \citenamefont {Johansson}(2007)}]{MarstorpJohansson_pof2007}%
  \BibitemOpen
  \bibfield  {author} {\bibinfo {author} {\bibfnamefont {L.}~\bibnamefont
  {Marstorp}}, \bibinfo {author} {\bibfnamefont {G.}~\bibnamefont
  {Brethouwer}}, \ and\ \bibinfo {author} {\bibfnamefont {A.~V.}\ \bibnamefont
  {Johansson}},\ }\bibfield  {title} {\enquote {\bibinfo {title} {A stochastic
  subgrid model with application to turbulent flow and scalar mixing},}\
  }\href@noop {} {\bibfield  {journal} {\bibinfo  {journal} {Phys.~Fluids}\
  }\textbf {\bibinfo {volume} {19}},\ \bibinfo {pages} {035107} (\bibinfo
  {year} {2007})}\BibitemShut {NoStop}%
\bibitem [{\citenamefont {Zidikheri}\ and\ \citenamefont
  {Frederiksen}(2010)}]{Frederiksen_ptrsa2010}%
  \BibitemOpen
  \bibfield  {author} {\bibinfo {author} {\bibfnamefont {M.}~\bibnamefont
  {Zidikheri}}\ and\ \bibinfo {author} {\bibfnamefont {A.~S.}\ \bibnamefont
  {Frederiksen}},\ }\bibfield  {title} {\enquote {\bibinfo {title} {Stochastic
  subgrid-scale modelling for non-equilibrium geophysical fluids},}\
  }\href@noop {} {\bibfield  {journal} {\bibinfo  {journal} {Phil. Trans. R.
  Soc. A}\ }\textbf {\bibinfo {volume} {368}},\ \bibinfo {pages} {145--160}
  (\bibinfo {year} {2010})}\BibitemShut {NoStop}%
\bibitem [{\citenamefont {Rasam}, \citenamefont {Brethouwer},\ and\
  \citenamefont {Johansson}(2014)}]{RasamJohansson_pof2014}%
  \BibitemOpen
  \bibfield  {author} {\bibinfo {author} {\bibfnamefont {A.}~\bibnamefont
  {Rasam}}, \bibinfo {author} {\bibfnamefont {G.}~\bibnamefont {Brethouwer}}, \
  and\ \bibinfo {author} {\bibfnamefont {A.~V.}\ \bibnamefont {Johansson}},\
  }\bibfield  {title} {\enquote {\bibinfo {title} {A stochastic extension of
  the explicit algebraic subgrid-scale models},}\ }\href@noop {} {\bibfield
  {journal} {\bibinfo  {journal} {Phys.~Fluids}\ }\textbf {\bibinfo {volume}
  {26}},\ \bibinfo {pages} {055113} (\bibinfo {year} {2014})}\BibitemShut
  {NoStop}%
\bibitem [{\citenamefont {Pope}(2014)}]{Pope_jfm2014}%
  \BibitemOpen
  \bibfield  {author} {\bibinfo {author} {\bibfnamefont {S.~B.}\ \bibnamefont
  {Pope}},\ }\bibfield  {title} {\enquote {\bibinfo {title} {The determination
  of turbulence-model statistics from the velocity-acceleration correlation},}\
  }\href@noop {} {\bibfield  {journal} {\bibinfo  {journal} {J.~Fluid~Mech.}\
  }\textbf {\bibinfo {volume} {757}},\ \bibinfo {pages} {R1 1--9} (\bibinfo
  {year} {2014})}\BibitemShut {NoStop}%
\end{thebibliography}%

\end{document}